\documentclass[epj]{svjour}
\usepackage{graphicx}
\usepackage{cite}
\usepackage{amssymb,amsmath}
\usepackage{flushend}
\begin{document}
\title{Flow probe of symmetry energy in relativistic heavy-ion reactions}
\author{
P.~Russotto,\inst{1}
M.D.~Cozma,\inst{2}
A.~Le F\`{e}vre,\inst{3}
Y.~Leifels,\inst{3}
R.~Lemmon,\inst{4}
Q.~Li,\inst{5}
J.~{\L}ukasik,\inst{6}
\and W.~Trautmann\inst{3}}
%
%\offprints{}         % Insert a name or remove this line
%
\institute{INFN-Sezione di Catania, I-95123 Catania, Italy
\and IFIN-HH, Reactorului 30, 077125 M\v{a}gurele-Bucharest, Romania
\and GSI Helmholtzzentrum f\"{u}r Schwerionenforschung GmbH, 
D-64291 Darmstadt, Germany 
\and STFC Daresbury Laboratory, Warrington WA4 4AD, United Kingdom
\and School of Science, Huzhou Teachers College, Huzhou 313000, P.R. China
\and IFJ-PAN, Pl-31342 Krak{\'o}w, Poland}
\date{Received: \today}
\abstract{
Flow observables in heavy-ion reactions at incident energies up to about 1 GeV per
nucleon have been shown to be very useful for investigating the reaction dynamics 
and for determining the parameters of reaction models based on transport theory.
In particular, the elliptic flow in collisions of neutron-rich heavy-ion systems 
emerges as an observable sensitive to the strength of the symmetry energy 
at supra-saturation densities. The comparison of ratios or differences of neutron 
and proton flows or neutron and hydrogen flows with predictions of transport models 
favors an approximately
linear density dependence, consistent with ab-initio nuclear-matter theories.
Extensive parameter searches have shown that the model dependence is comparable to 
the uncertainties of existing experimental data. Comprehensive new flow data of high 
accuracy, partly also through providing stronger constraints on model parameters, 
can thus be expected to improve our knowledge of the equation of state of asymmetric 
nuclear matter. 
 % end of PACS codes
} %end of abstract
\authorrunning{P. Russotto et al.}
\titlerunning{Flow probe of symmetry energy}
\maketitle
\section{Introduction}
\label{sec:intro}
Heavy-ion reactions at relativistic energies and small impact parameters proceed through 
violent initial stages during which highly excited and compressed nuclear matter is 
temporarily produced. 
The study of these reactions is thus a means of gaining information on extreme-matter 
properties which recently has advanced to the tera-electron-volt regime at the 
Large Hadron Collider at the CERN laboratory~\cite{abelev07,aamodt10,adare12,aad12}. 

\begin{figure}[!htb]
 \leavevmode
 \begin{center}
  \includegraphics[width=0.90\columnwidth]{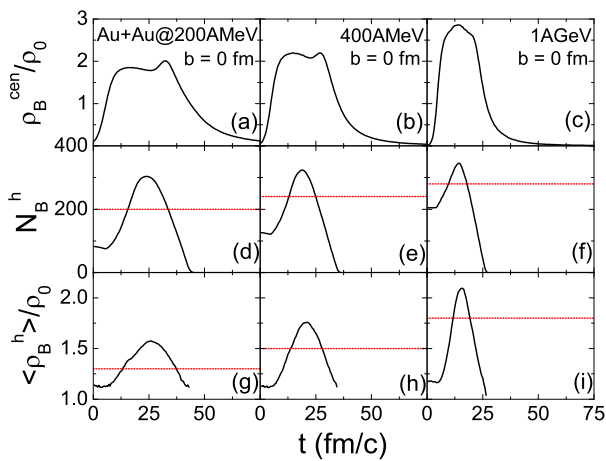}
 \end{center}
\vskip -0.1cm  
  \caption{Temporal evolution of central baryon density (top), 
baryon multiplicity (middle), and average baryon density (bottom) 
in the high-density ($\rho > \rho_0$) phase of central $^{197}$Au + $^{197}$Au collisions 
at three energies ($\rho_0$ is the saturation density). 
Horizontal lines indicate the average values used in the thermal 
model of Ref.~\protect\cite{xu13} (reprinted with permission from 
Ref.~\protect\cite{xu13}; 
Copyright (2013) by the American Physical Society).}

\label{fig:xu_density}
\end{figure}

The present review is focussed on the lower end of the relativistic domain with 
incident energies between several hundred MeV up to about one GeV per nucleon. Here, the 
central densities may reach values of up to two or three times the saturation value
while the dynamics is still dominated by hadronic degrees of freedom. This is illustrated 
in Fig.~\ref{fig:xu_density} with results from isospin-dependent Boltzmann-Uehling-Uhlenbeck 
(IBUU) transport~model calculations~\cite{xu13}. The high-density phase is short, typically 
20 fm/c, but involves up to three quarters of the total baryon multiplicity of the system 
(Fig.~\ref{fig:xu_density}, middle row of panels). 
Nuclear matter at densities of this order is believed to be present in 
neutron stars and also temporarily formed during the core-collapse phase of supernova 
explosions. Modeling these astrophysical phenomena requires the knowledge of nuclear 
matter properties far away from saturation and far away from symmetry. 
The equation-of-state (EoS) of neutron-rich asymmetric matter has, therefore, received 
particular attention recently, motivated by the impressive progress made in astrophysical 
observations and measurements~\cite{lattprak07,lipr08,ditoro10,demorest}, and especially also
at high density where its behavior is least well known~\cite{lipr08,ditoro10,fuchs06}.  

\vskip -0.2cm 

Colliding heavy ions at relativistic energies represents a possible means 
for studying the nuclear equation of state at supra-saturation densities
in laboratory experiments. 
It still remains a challenge, however, to find observables suitable for extracting 
information on its properties during the brief compression phase~\cite{dani00,dani02}.
This has been an intense field of research during the past decades which has rapidly concentrated
on collective flows and meson production. 
The pressure gradients due to compression and the increased collision rates 
at high density are expected to affect these observables. 
Studies of flow and kaon production within the framework of transport theory
have indeed both been essential for reaching the present consensus that a soft EoS with
compressibility $K \approx 230$~MeV and momentum dependent interactions best describes the
response of symmetric nuclear matter to compression~\cite{dani02,sturm01,fuchs01,hartnack06}.

The same observables appear naturally as primary candidates for investigating the equation
of state of asymmetric matter, the so-called asy-EoS. It is usually expressed in the
form of the symmetry energy, the subject of this topical volume,
which is the difference between the energies per nucleon of neutron matter and of symmetric 
matter. It represents the response of nuclear matter to asymmetry. Its study, consequently, 
requires differential observables measuring differences as a function of asymmetry or isotopic 
pairs of observables that respond differently to a compression of neutron-rich matter.
As the symmetry energy appears in nearly every aspect of nuclear structure and reactions, 
a wide variety of possibilities exists. Many different kinds of constraints have been 
identified and quantified for the density regime below saturation, down to extremely low 
densities. The accompanying articles of this topical 
volume present a detailed picture of the various studies made and of the convergence 
achieved in recent years (see also, e.g., Refs.~\cite{tsang12,moeller12,chen12}).

In the density regime exceeding saturation, the symmetry energy is still largely unknown 
for several reasons. Phenomenological forces are well constrained near or just below
saturation but lead to largely diverging results if they are extrapolated to higher 
densities~\cite{fuchs06,brown00}.
Microscopic many-body calculations with realistic potentials face the difficulty
that three-body forces and short-range correlations are not sufficiently well known
at higher densities at which their importance increases~\cite{subedi08,xuli10,steiner12}. 
Chiral effective field theories have no free parameters for three-body forces, when 
applied to systems with only neutrons, but require extrapolations for reaching densities 
beyond saturation~\cite{hebeler10prc}. 
Even the magnitude of the kinetic contribution,
related to the nuclear Fermi motion and considered as principally understood, could 
possibly be modified by a redistribution of nucleon momenta due to short-range correlations in 
high-density nuclear matter~\cite{carb11,alvioli13}.
The need for experimental high-density probes is thus obvious, with the above-mentioned 
collective flows and sub-threshold particle production as the main candidates.

\section{High-density probes}
\label{sec:probes}

Collective flows in nucleus-nucleus collisions, appearing in the form of anisotropic
particle emissions as a result of collective velocity fields, have been studied rather
intensely for many years
(for reviews see, e.g.,~\cite{reisdorf97,herrmann99}). 
Measurements at beam energies up to  several GeV per nucleon were used to extract 
the EoS of nuclear matter from quantitative comparisons with the results of microscopic
transport calculations~\cite{dani00,dani02,stoecker86,reisdorf12}.
Significant progress has been made. 

\begin{figure}[!htb]
 \leavevmode
 \begin{center}
  \includegraphics[angle=270,width=0.80\columnwidth]{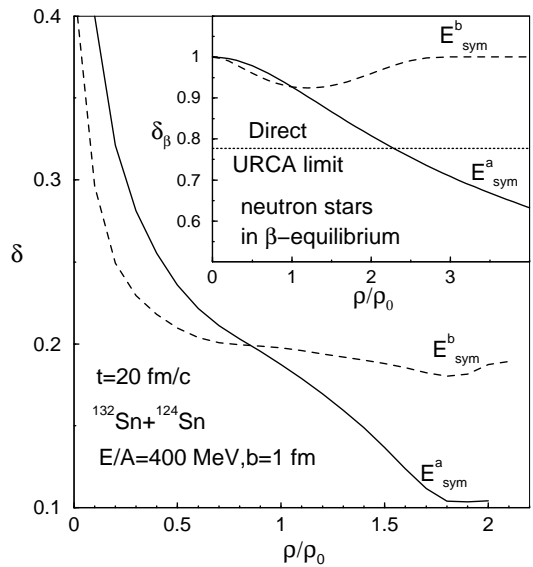}
 \end{center}
\vskip -0.1cm  
  \caption{Isospin asymmetry $\delta = (\rho_n -\rho_p)/\rho$ as a function of the 
normalized density $\rho/\rho_0 $ 
at time $t = 20$~fm/$c$ in the $^{132}$Sn + $^{124}$Sn reaction with 
a stiff ($E^a_{\rm sym}$) and with a soft ($E^b_{\rm sym}$) density dependence of the 
nuclear symmetry energy. 
The corresponding correlation for neutron stars in $\beta$-equilibrium is shown in
the inset (reprinted with permission from Ref.~\protect\cite{li02}; 
Copyright (2002) by the American Physical Society).}

\label{fig:li_prl02fig2}
\end{figure}
%\vskip -0.3cm 

The use of differential isotopic flows for the study of the asy-EoS has been suggested
by Bao-An Li whose calculations with an isospin-dependent hadronic transport model indicated 
parallels in the density-dependent isotopic compositions of neutron stars and of the transient 
systems formed in collisions of neutron-rich nuclei. They exhibited a similar dependence on 
the high-density
behavior of the nuclear symmetry energy used in the calculations~\cite{li02}. 
This is illustrated in Fig.~\ref{fig:li_prl02fig2} which points to the remarkable possibility
of gaining information on macroscopic astrophysical objects from laboratory experiments with
atomic nuclei smaller by 55 orders of magnitude in mass.

The comparatively small asymmetry of available nuclei represents a major difficulty,
however. There will be an important role to be played by radioactive secondary beams, 
as assumed in the example shown in Fig.~\ref{fig:li_prl02fig2}, 
but the symmetry effects will still be small relative to those of the dominating isoscalar 
forces. It is thus essential to identify differential observables as, e.g., differences or 
ratios of observables measured for isotopic partner systems or reactions. 
One hopes to enhance the response to asymmetry as the isoscalar dynamics largely cancels.
The so-called differential directed flow proposed by Li~\cite{li02} is the
difference of the multiplicity-weighted directed flows of neutrons and protons. Directed
flow describes the rapidity dependence of the mean in-plane transverse momenta of observed 
reaction products. Isotopic yield ratios or double ratios represent another possible class 
of observables. The double neutron-to-proton ratios obtained from isospin-asymmetric but 
mass-symmetric pairs of reactions proposed in Refs.~\cite{lili06,feng12} are being successfully used
at lower energies~\cite{famiano06}.

Besides the transverse directed flow, also the elliptic flow has been studied with model 
calculations to test its usefulness as a probe of the 
asy-EoS~\cite{lipr08,scalone99,lisustich01,greco03,baran05}. 
Elliptic flow relates to the azimuthal anisotropy of particle emissions, mainly differentiating
between predominantly in-plane emissions as recently observed in ultrarelativistic 
heavy-ion collisions~\cite{abelev07,aamodt10,adare12,aad12} and the out-of-plane emissions 
or squeeze-out observed in the present regime of lower energies as a
consequence of the pressure build-up in the collision zone~\cite{gutbrod90}.  
A particular encouragement was provided by transport calculations with quantum-molecular
dynamics (QMD) models
for $^{197}$Au + $^{197}$Au collisions at 400 MeV per nucleon according to which the elliptic 
flow of free neutrons and protons responds significantly differently
to variations of the parameterization of the symmetry energy~\cite{russotto11,cozma11}.

A data set to test these predictions has been available from earlier experiments of the 
FOPI/LAND Collaboration. It was originally collected and shown to provide evidence for the  
squeeze-out of neutrons emitted in $^{197}$Au + $^{197}$Au collisions at 400 MeV per
nucleon~\cite{leif93,lamb94}. The capability of the Large Area Neutron Detector 
LAND~\cite{LAND} used in these experiments
of detecting neutrons as well as charged particles permitted the differential analysis of the
observed flow patterns in the form of flow ratios~\cite{russotto11} or flow 
differences~\cite{cozma11}. Both analyses favor a density dependence between moderately soft 
and moderately stiff, close to the predictions of ab-initio calculations using realistic forces
(see below) and also consistent with the observations made at sub-saturation 
density (see, e.g., Refs.~\cite{tsang12,moeller12,chen12} and pertinent articles in this topical volume).

Whether meson production yields will become similarly useful for the same purpose is not so 
clear at present. It has been suggested that the ratio of the anti-strange kaon isospin 
partners, K$^+$/K$^0$, may serve as a useful observable~\cite{qli05,ferini06}. 
The predicted effects were not very large but the production of kaons through $\Delta$
resonances and their weak interaction with the nuclear medium 
distinguish them as direct messengers from the high-density zone.
Subthreshold kaon production for the isotopic pair of reactions  $^{96}$Zr + $^{96}$Zr 
and $^{96}$Ru + $^{96}$Ru was studied by the FOPI Collaboration at 1.53 GeV per 
nucleon~\cite{xlopez07}.
The K$^+$ mesons were identified by correlating their momentum determined with the central 
drift chamber and their velocity measured with the time-of-flight barrel of the FOPI detector. 
The K$^0_s$ mesons were identified via their weak decay into $\pi^+$ and $\pi^-$ and
by reconstructing the displaced decay vertices in the central drift chamber. Because of the 
resulting significantly different detection efficiencies, a double ratio was formed from 
the K$^+$/K$^0$ production ratios obtained for the two collision systems and compared to 
model calculations. A significant sensitivity of this observable to the chosen 
stiffness of the asy-EOS was expected from the calculations for infinite nuclear matter. It
diminished, however, by one order of magnitude when the calculations were performed for 
the actual heavy-ion collisions studied in the experiment~\cite{xlopez07}. The measured 
double ratio is satisfactorily reproduced irrespective of the choice made for the 
asy-EoS.

An even more puzzling situation is  
encountered in the case of the $\pi^-/\pi^+$ yield ratios measured by the FOPI
Collaboration at several energies up to 1.5 GeV per nucleon and for the four 
mass-symmetric systems  $^{40}$Ca + $^{40}$Ca, $^{96}$Zr + $^{96}$Zr,  $^{96}$Ru + $^{96}$Ru,
and $^{197}$Au + $^{197}$Au~\cite{reisdorf07}.
Theoretical analyses of this data set came to
rather conflicting conclusions, suggesting everything from a rather stiff to a
super-soft behavior of the symmetry energy~\cite{xiao09,feng10,xie13}. 
Among them, the super-soft result has initiated a broad discussion of how it might be 
reconciled with observed properties of neutron stars~\cite{xiao09,baoan11,wen09}. 

\begin{figure}[!htb]
 \leavevmode
 \begin{center}
  \includegraphics[width=0.9\columnwidth]{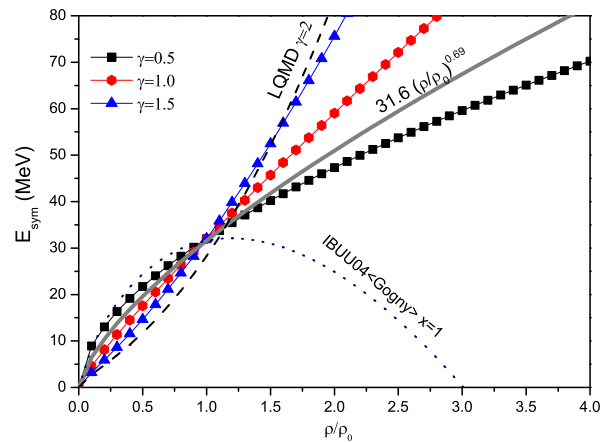}
 \end{center}
\vskip -0.1cm  
  \caption{Parameterizations of the nuclear symmetry energy as used in transport codes: 
three parameterizations of the potential term used in the 
UrQMD (Ref.~\protect\cite{qli05}) 
with power law coefficients $\gamma = 0.5, 1.0$, and 1.5 (lines with symbols as indicated), 
the result with $\gamma = 0.69$ obtained from analyzing
isospin diffusion data with the IBUU04 (full line, Ref.~\protect\cite{lichen05}),
and the super-soft and stiff parameterizations obtained from analyzing the $\pi^-/\pi^+$ 
production ratios with the IBUU04 (dotted line, Ref.~\protect\cite{xiao09}) and the ImIQMD 
(dashed line labeled LQMD, Ref.~\protect\cite{feng10}) transport models
(from Ref.~\protect\cite{guo12},
reprinted with kind permission from Springer Science+Business Media).
}
\label{fig:params}
\end{figure}

Calculations with standard parameters fell below the measured $\pi^-/\pi^+$ yield ratios. 
Therefore, more extreme assumptions had to be made to reach the experimental values.
In the analysis of Xiao et {\em et al.}~\cite{xiao09} with the IBUU04 transport model
supplemented with the Gogny-inspired momentum-dependent parameterization of the
symmetry energy~\cite{das03}, 
the measured dependences on the collision system and impact parameter have been well 
reproduced but
only by assuming a super-soft density-dependence of the symmetry energy, close to
the $x=+1$ case shown in Fig.~\ref{fig:params}. 
The analysis of Feng and Jin~\cite{feng10} carried out with the improved isospin-dependent
QMD (ImIQMD) and a power-law 
parameterization of the potential part of the symmetry energy yielded an equally satisfactory
description of the data with the stiff choice $\gamma = 2$ for the
power law exponent (cf. Fig.~\ref{fig:params}). 
Most recently, Xie {\em et al.} addressed the same issue within the Boltzmann-Langevin approach
and a power-law parameterization and obtained again support for a super-soft scenario for 
the symmetry energy~\cite{xie13}.

This situation is clearly unfortunate because the expected variations themselves,
of up to 20\% for soft versus stiff parameterizations, are rather large. Possible reasons
for it may lie in the treatment of the $\Delta$ dynamics in transport models and in
competing effects of the mean fields and $\Delta$ thresholds whose weights may be varying
among the different approaches~\cite{ditoro10,ferini05,wolter12}.
The role of partial cancellations of $s$-wave and $p$-wave effects in the nuclear medium
which causes a reduction of the $\pi^-/\pi^+$ ratios has recently been pointed out
by Xu {\em  et al.}~\cite{xu13}. Including the isospin-dependent pion in-medium effects
is thus important, even though not trivial in transport models. 
For a more detailed discussion and interpretation, the reader is referred to the 
article by Zhi-Gang Xiao {\em  et al.} in this topical volume. 

The conflicting interpretations of the meson data
emphasize the need for reaching an improved understanding of the reaction mechanisms 
in order to achieve more robust results. Possible model dependences are an urgent issue
also in the case of the differential elliptic flow and the conclusions regarding the 
density dependence of the symmetry energy obtained from there.
Both interpretations of the FOPI/LAND data have addressed this question and demonstrated
that the chosen differential observables are fairly stable with respect to variations of
global parameters of the calculations as, e.g., the isoscalar EoS or the parameterization
of the nucleon-nucleon cross sections~\cite{russotto11,cozma11}. A more comprehensive study
of the effects of global parameter variations within their presently known limits has
recently been completed and will be presented in the final section. 
It confirms the earlier results by showing that the overall model dependence is small enough 
to prevent it from concealing the sensitivity 
of the differential flows to the asy-EoS~\cite{cozma13}. 
However, narrower constraints for the global parameters will be 
necessary if a more precise determination is to be achieved.

An urgent need to improve the statistical accuracy beyond that of the existing data set has
equally become obvious from the studies performed on the FOPI/LAND data. It has initiated a 
dedicated measurement of collective flows in collisions of $^{197}$Au + $^{197}$Au as well 
as of the $^{96}$Zr + $^{96}$Zr and $^{96}$Ru + $^{96}$Ru pair of systems carried out in 
2011 at the GSI laboratory~\cite{s394,s394_nn2012}. The LAND~\cite{LAND} detector has been operated
together with a subset of the CHIMERA~\cite{chimera} detector array complemented with additional 
detector systems aiding in the measurement of the reaction plane orientation and of the flow of 
light fragments. Details of the experiment and the present status of the analysis will 
be given further below after the elliptic flow and its interpretation with transport models
have been reviewed. A brief introductory overview addressing the relevance of elliptic flow 
in the study of the symmetry energy at supra-saturation density is available in 
Ref.~\cite{traut12}.

\section{Symmetry energy and parameterization}

The symmetry energy is familiar from the study of atomic masses whose dependence 
on their isotopic nuclear composition is accounted for with the symmetry term in the 
Bethe-Weizs\"{a}cker mass formula. A density dependence is already suggested by the use of 
individual bulk and surface terms in more refined mass-formulae. Their values are, e.g.,
28.1 MeV and 33.2 MeV, respectively, in the well-known parameterization of Myers and 
Swiatecki~\cite{ms}.
In the Fermi-gas model, the density dependence is given by a proportionality
to $(\rho / \rho_0)^{\gamma}$ with an exponent $\gamma = 2/3$,
where $\rho_0 \approx 0.16~{\rm nucleons/fm}^{3}$ is the saturation density.
The coefficient of this so-called kinetic contribution to the symmetry energy is 
$\epsilon_F/3$, where $\epsilon_F \approx 28$~MeV is the Fermi energy. It
amounts to only about 1/3 of the symmetry term of $\approx 30$~MeV for nuclear matter at 
saturation. The major contribution
is given by the potential term reflecting properties of the nuclear forces.

\begin{figure}[!htb]
 \leavevmode
 \begin{center}
  \includegraphics[width=0.95\columnwidth]{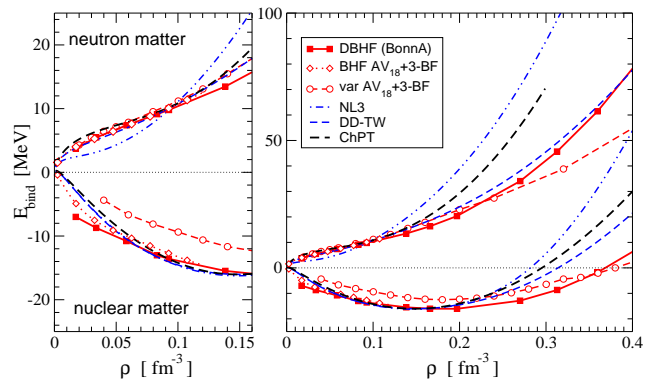}
 \end{center}
\vskip -0.1cm  
  \caption{EoS in nuclear matter and neutron matter.  
BHF/DBHF and variational calculations with realistic forces are compared to  
phenomenological density functionals NL3 and DD-TW and to 
ChPT. The left panel zooms the low density range  
(from Ref.~\protect\cite{fuchs06}, 
reprinted with kind permission from Springer Science+Business Media).
}
\label{fig:fuchs06}
\end{figure}

\begin{figure}[!htb]
 \leavevmode
 \begin{center}
  \includegraphics[width=0.95\columnwidth]{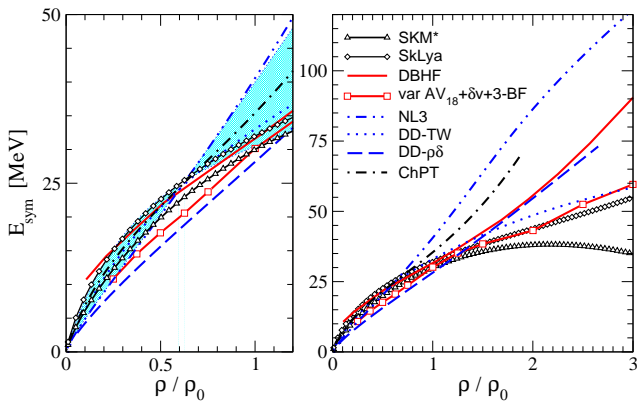}
 \end{center}
\vskip -0.1cm  
  \caption{Symmetry energy as a function of density as predicted by different  
models. The left panel zooms the low density range up to saturation. The full lines
represent the DBHF and variational approaches using realistic forces   
(from Ref.~\protect\cite{fuchs06}, 
reprinted with kind permission from Springer Science+Business Media).
}
\label{fig:fuchs06b}
\end{figure}

Microscopic many-body calculations have presented us with a variety of predictions for the nuclear
equation of state~\cite{fuchs06,dani02,baldo04}. 
The examples shown in Fig.~\ref{fig:fuchs06} for the two cases of 
symmetric nuclear matter and of pure neutron matter demonstrate that, overall, the results
are quite compatible among each other, except for densities exceeding saturation for 
which the predictions diverge. 
The symmetry energy $E_{\rm sym}$ is defined as the coefficient of the quadratic 
term in an expansion of the energy per particle in the asymmetry  
$\delta = (\rho_n-\rho_p)/\rho$, where $\rho_n, \rho_p,$ and $\rho$
represent the neutron, proton, and total densities, respectively,
\begin{equation}
E/A(\rho,\delta) = E/A(\rho,\delta = 0) + E_{\rm sym}(\rho)\cdot \delta^2 + \mathcal{O}(\delta^4).
\label{eq:e_sym}
\end{equation}

\noindent In the usual quadratic approximation, the symmetry energy is the difference 
between the energies of neutron matter ($\delta = 1$) and symmetric matter ($\delta = 0$).
As expected from Fig.~\ref{fig:fuchs06}, also the predictions for the symmetry energy diverge at high densities
while most of them coincide near or slightly below saturation, 
the density range at which constraints from finite nuclei are valid (Fig.~\ref{fig:fuchs06b}).

In calculations using realistic forces fitted to two- and three-nucleon data, the 
uncertainty is mainly related to the short-range behavior of the nucleon-nucleon force
and, in particular, to the three-body and tensor forces~\cite{subedi08,xuli10,steiner12}. 
The three-body force has been shown to make an essential contribution of several MeV 
to the masses of light nuclei~\cite{wiringa02}. 
The extrapolation to higher densities of the partly phenomenological terms used there is, 
however, highly uncertain~\cite{xuli10}. 
The general effect of including three-body forces in the calculations is a 
stiffening of the symmetry energy with increasing density~\cite{zhli06,burgio08,hebe10}.
Short-range correlations become also increasingly important at higher densities;
results from very recent new experiments will, therefore, have a 
strong impact on predictions for high-density nuclear matter~\cite{subedi08,carb11,alvioli13}.

It is also well known that nuclear mean fields are momentum 
dependent~\cite{lipr08,ditoro10,giordano10,feng12}. 
It is evident, e.g., in the energy dependence of the nuclear optical potential. 
The dominating effect is seen in the isoscalar sector but the isovector momentum
dependence may also be important. It manifests itself as an energy dependence of the 
isospin-dependent part of the optical potential
but can also be expressed in terms of a difference of the effective masses of protons and
neutrons~\cite{lipr08,ditoro10,fuchs06}. Even the ordering of these effective masses is still an open
problem. It has, moreover, been shown that the 
effective mass differences and the asymmetry dependence of the EoS are both influencing
particle yields and flow observables, and that additional observables will be needed
to resolve the resulting ambiguity~\cite{ditoro10,giordano10,feng12}.
%\vskip -0.2cm 

Transport theory provides the tools for following the temporal evolution of nuclear reactions.
For describing the composition-dependent part of the nuclear mean field, parameterizations
based on potential models are commonly used. In the ultrarelativistic QMD (UrQMD) model  
of the group of Li and Bleicher~\cite{qli05,qli06,Li:2006ez},
the potential part of the symmetry energy is defined with two parameters, 
the value at saturation density, usually taken as 22 MeV in their calculations, 
and the power-law coefficient $\gamma$ describing the dependence on
density as $(\rho / \rho_0)^{\gamma}$.
In the QMD model of the T\"{u}bingen group~\cite{cozma11,kho92,uma98} (both approaches are 
discussed in more detail in Section~\ref{sec:qmd}), the nuclear potential of Das {\it et al.} 
with explicit momentum dependence in the isovector sector is used 
(MDI interaction, Ref~\cite{das03}). 
There, as in the IBUU04 developed by the groups of Li and Chen~\cite{lipr08,lichen05}, 
the density dependence of the symmetry energy is characterized by a parameter $x$ 
appearing in the potential expressions.
Examples of these parameterizations and of results obtained from the analysis of experimental
reaction data are given in Fig.~\ref{fig:params}. The stiff ($E^a_{\rm sym}$) and 
soft ($E^b_{\rm sym}$) density dependences of Fig.~\ref{fig:li_prl02fig2} correspond 
approximately to the cases $\gamma = 1$ and $x = 1$ shown there.

Parameterizations of this kind have the consequence that, once the symmetry energy at the 
saturation point is fixed, a single value at a different density or, alternatively, 
the slope or curvature at any density will completely determine the 
parameterization. Measurements of a variety of observables in nuclear structure and reactions 
have been used in this way to obtain results for the density dependence of the 
symmetry energy. They are often
expressed in the form of the parameter $L$ which is proportional to the slope 
at saturation, 
\begin{equation}
L = 3\rho_0 \cdot dE_{\rm sym}/d\rho |\rho_0.
\label{eq:l}
\end{equation} 

\noindent Most results with their errors fall into the interval 20~MeV $\le L \le$ 100~MeV 
and are compatible with a most probable value $L\approx 60$~MeV, roughly 
corresponding to a power-law coefficient 
$\gamma = 0.6$~\cite{lipr08,tsang12,moeller12,tsang09,baoan11,lattlim12}. 
The full line in Fig.~\ref{fig:params} represents, e.g., the result $\gamma = 0.69$ 
deduced by Li and Chen from the MSU isospin-diffusion data and the neutron-skin thickness 
in $^{208}$Pb~\cite{lichen05}.
The corresponding slope parameter is $L=65$~MeV. 
Rather similar constraints have been deduced from very recent investigations and observations 
of neutron-star properties~\cite{steiner12,hebe10,steiner10,steiner13,baoan12}. 
Carefully constrained microscopic calculations with realistic potentials have also been 
shown to be fully compatible with these results as, e.g., $L = 66.5$~MeV obtained by 
Vida\~{n}a {\em  et al.}~\cite{vidana09}. The present situation here 
as well as the information on the density dependence of the symmetry energy deduced
from nuclear structure and reactions probing nuclear matter near and below saturation are
given in the accompanying articles of this topical volume. 

High expectations are placed on
the determination of the neutron-skin thickness of $^{208}$Pb and other
neutron-rich nuclei by measuring the parity-violating contribution to electron scattering 
at high energy~\cite{abrah12}. It will offer a practically model-free access to the slope 
at saturation, even though it is obtained by probing  
nuclear matter at an average density of typically 2/3 of this value~\cite{horo12}.
In the more distant future, gravitational wave detection may reach a sensitivity permitting
the study of tidal deformations of neutron stars in coalescing binary systems, considered
as strongly depending on the high-density symmetry energy~\cite{baoan12}.

\section{Directed and elliptic flows}

Collective flows are known to be sensitive to essential features of the reaction dynamics 
since many years. 
Early studies have concentrated on the transition from mean-field 
dynamics to nucleon-nucleon collision dominated dynamics in the Fermi-energy 
domain~\cite{bertsch87,lemmon99}. In experiments not sensitive to the absolute sign of the
preferred direction of particle emission, it appeared as a so-called disappearance of 
flow, referring here to directed flow~\cite{westfall98}. 
The corresponding energy, also called balance energy and observed at
roughly 50 to 100 MeV per nucleon, and its dependence on the mass and on the isotopic 
compositions of the colliding nuclei has attracted 
considerable interest~\cite{westfall93,pak97,pak97a,zhang06}.
Directed flow data, measured at even lower energies for selected isotopic and isobaric 
pairs of collision systems, have more recently been shown to provide information on the
symmetry energy and its density dependence (Ref.~\cite{kohley10,kohley12} and accompanying 
article by Kohley and Yennello).

The influence of the density-dependent symmetry energy on the balance energy 
of heavy collision systems from $^{96}$Zr + $^{96}$Zr to $^{197}$Au + $^{197}$Au 
has also been studied with the updated version of the UrQMD transport model~\cite{guo12}.
It was, e.g., found that the balance energy of neutrons is particularly sensitive to the density 
dependence of the symmetry potential energy, an encouragement for new experiments since
data of this kind do not exist up to now. In another systematic UrQMD study, the sign and 
magnitude of directed flow
were shown to depend crucially on several of the many parameters entering the 
calculation~\cite{qli11}. To correctly reproduce the magnitude of directed $Z=1$ flow 
observed for semi-central $^{197}$Au + $^{197}$Au collisions between 40 and 150 MeV per nucleon,
a careful adjustment of the isoscalar EoS and of the density and momentum dependence of
the in-medium nucleon-nucleon elastic cross sections had to be made. It emphasizes the need
for consistent considerations of both, the mean field and the two-body collisions in
transport models.

Reactions near the balance energy do not strongly compress the colliding matter.
As shown above (Fig.~\ref{fig:xu_density}), several hundred MeV per nucleon up to 
$\approx 1$~GeV per nucleon are needed to reach densities of two to three times the saturation 
density in central collisions~\cite{xu13,li_npa02}. 
At these energies, the pressure produced during the short high-density intervals initiates a collective outward 
motion whose strength will depend on the equation of state. The resulting direct and elliptic 
flows have been confronted with EoS model predictions and extreme assumptions have been shown to
be ruled out~\cite{dani02}. 

\begin{figure}[!htb]
 \leavevmode
 \begin{center}
  \includegraphics[width=0.85\columnwidth]{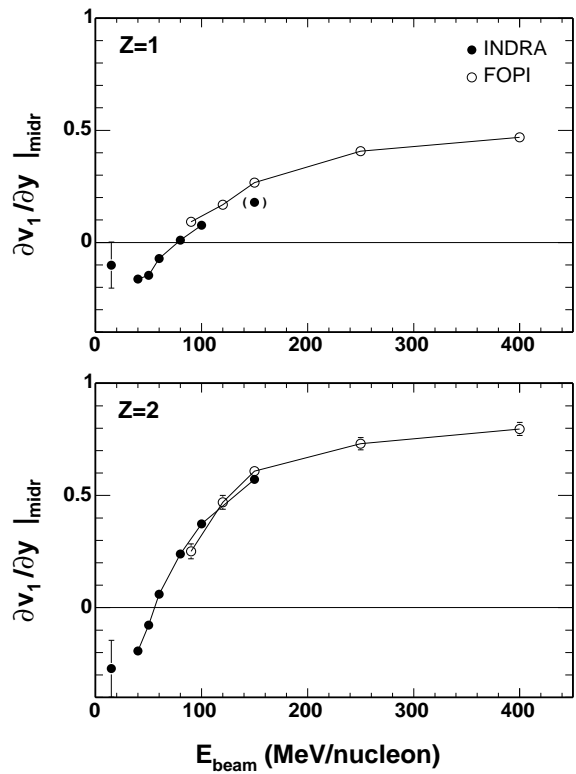}
 \end{center}
\vskip -0.1cm  
\caption{Slopes of directed flow $\partial v_{1}/\partial y$ for $Z=1$ (top)
  and $Z=2$ (bottom) particles integrated  over transverse momentum $p_t$ for mid-central
  $^{197}$Au + $^{197}$Au
  collisions (2--5.5 fm). The open and filled symbols represent the 
  FOPI~\protect\cite{andronic01} and the INDRA data, respectively. The uncertainty
  at $15$~MeV per nucleon is mainly statistical. The INDRA point, in brackets, at 
  $150$~MeV per nucleon in the top panel is biased due to experimental inefficiencies for $Z=1$
  at this energy
(from Ref.~\protect\cite{andronic06},
reprinted with kind permission from Springer Science+Business Media).
}
\label{fig:v1corr}
\end{figure}

Dynamical flow observables, expected to be influenced by the symmetry energy in asymmetric 
systems, have been proposed by several groups as probes for the equation of state at high 
density~\cite{li02,scalone99,lisustich01,greco03,baran05}. To be sensitive to the symmetry part, 
differential observables are required. The so-called differential neutron-proton 
flow is the difference of the parameters describing the collective motion of free 
neutrons and protons weighted by their numbers~\cite{li02}. According to the simulations,
this observable minimizes the influence of the isoscalar part in the EoS while maximizing 
that of the symmetry term~\cite{yong06}. Its proportionality to the particle
multiplicities, however, makes its determination very dependent on the experimental 
efficiencies of particle detection and identification and on the precise procedure for 
distinguishing free and bound nucleons in calculations. Therefore, also  
differences or ratios of directed and elliptic flows have been considered. 

It has become customary to express both, directed and elliptic flows, 
and possibly also higher flow components by means of a Fourier decomposition
of the azimuthal distributions measured with respect to the orientation of the 
reaction plane $\phi_R$~\cite{voloshin96,ollitrault97,poskanzer98},   
\begin{equation} 
\frac{dN}{d(\phi-\phi_R)} = \frac{N_0}{2\pi} 
\left( 1+2 \sum_{n\geq1} v_n \cos n(\phi-\phi_R)\right),
\label{eq:defvn} 
\end{equation}

\noindent where $N_0$ is the azimuthally integrated yield. The
coefficients $v_{n} \equiv \langle\cos n(\phi-\phi_R)\rangle$ are functions of 
particle type, impact parameter, rapidity $y$, and the transverse momentum $p_t$. 

\begin{figure}[!htb]
 \leavevmode
 \begin{center}
  \includegraphics[width=0.90\columnwidth]{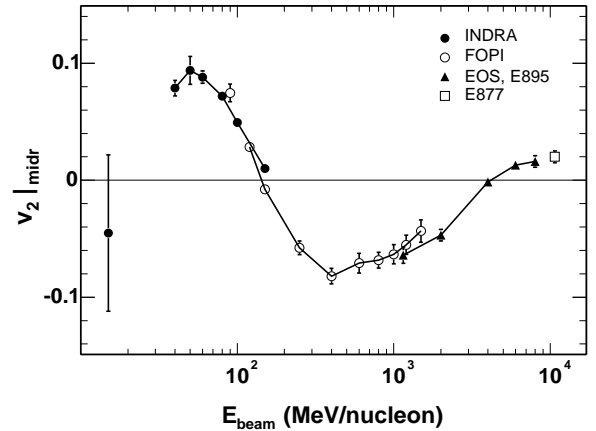}
 \end{center}
\vskip -0.1cm  
\caption{Elliptic flow parameter $v_{2}$ at mid-rapidity for $^{197}$Au + $^{197}$Au 
collisions at intermediate impact parameters (about 5.5-7.5 fm) as a function of incident
energy, in the beam frame. The filled and open circles represent the INDRA
and FOPI data~\protect\cite{lukasik05,andronic05}, respectively, for $Z=1$ particles,
the triangles represent the EOS and E895 data~\protect\cite{pinkenburg99} for
protons and the square represents the E877 data~\protect\cite{bmunzinger98}
for all charged particles
(from Ref.~\protect\cite{andronic06},
reprinted with kind permission from Springer Science+Business Media).
}
\label{fig:v2corr}
\end{figure}

Excitation functions of the directed and elliptic flows for light charged particles
from $^{197}$Au+$^{197}$Au collisions are shown in Figs.~\ref{fig:v1corr},~\ref{fig:v2corr}.
The directed flow is quantified as the derivative of $v_{1}$ with respect to rapidity, 
taken at mid-rapidity, and a positive slope, by definition, identifies a predominance
of repulsion. The data for $Z=1$ and $Z=2$ particles illustrate the $Z$ dependence of the
balance energy, being higher by more than 10 MeV per nucleon for the hydrogen isotopes
as compared to helium.

The excitation function of the elliptic flow, quantified as $v_{2}$ at midrapidity and
shown in Fig.~\ref{fig:v2corr} for $Z=1$ particles in $^{197}$Au + $^{197}$Au collisions 
from various experiments, extends up to the AGS regime of several GeV per nucleon incident 
energies at which in-plane flow starts to dominate again. 
Squeeze-out perpendicular to the reaction plane ($v_2 < 0$), 
as a result of shadowing by the spectator remnants is observed
at incident energies between about 150~MeV per nucleon and 4~GeV per nucleon with a maximum near
400 MeV per nucleon.  At lower energies, the collective rotation in the mean-field 
dominated dynamics causes the observed in-plane enhancement of emitted reaction 
products~\cite{tsang93}.
The two figures also illustrate the precision that can be reached in flow measurements.
The reliability of the applied methods is demonstrated by the good agreement of results 
from different experiments in the overlap regions of the studied intervals in collision 
energy~\cite{andronic06,lukasik05}. 

Elliptic flow has become an important observable at other energy regimes as well. At
ultrarelativistic energies, the observation of the constituent-quark scaling of elliptic flow
is one of the prime arguments for deconfinement during the early collision phase,
and properties of the formed quark-gluon liquid are deduced from the observed
magnitude of collective motions~\cite{abelev07,aamodt10,adare12,aad12,fries08,snell11}.
It implies that elliptic flow develops very early in the collision which is valid also
in the present range of relativistic energies as confirmed by calculations~\cite{dani00}.

Experimentally, the azimuthal angle of emission is determined with respect to the 
orientation of a reaction plane that has been reconstructed from observed emission patterns.
Several methods have been proposed (see, e.g., Ref.~\cite{andronic06}) which have 
in common an overall dependence of their accuracy on the emitted particle types and 
multiplicities of the considered reaction. The so-called dispersion of the reaction 
plane refers
to the uncertainty of the reconstructed azimuthal orientation. The deduced flow parameters
decrease in the case of poorly determined experimental reaction planes and corrections
are necessary for which, however, quite refined methods exist. Their magnitude increases 
with the order of the Fourier coefficient considered.

\begin{figure}[htb!]
\centering
\includegraphics*[width=0.80\columnwidth]{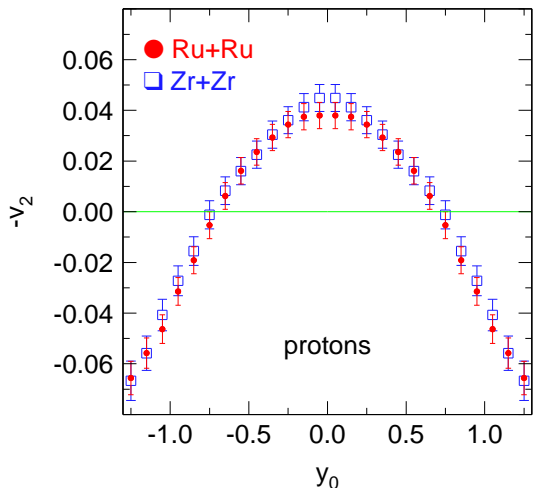}
\vskip -0.1cm
  \caption{Elliptic flow $-v_2 (y_0)$ of protons in $^{96}$Zr + $^{96}$Zr (blue open squares) and 
$^{96}$Ru + $^{96}$Ru (red dots) collisions for 1.5~GeV per nucleon incident energy and centrality
$0.25< b_0 < 0.45$. Note the inversion of $v_2$ and the definitions of the reduced impact
parameter $b_0 = b/b_{\rm max}$ and of the normalized rapidity in the c.m. frame, 
$y_0 = y/y_p$, with $y_p$ denoting the projectile rapidity
(reprinted from Ref.~\protect\cite{reisdorf12}, Copyright (2012), with permission from 
Elsevier).
}
\label{fig:fopi_ruzr}
\end{figure}    %fig. 8

In the INDRA and FOPI data sets included in the figures, the reaction plane has been 
reconstructed using the so-called Q-vector method in slightly different forms.
The Q-vector representing the orientation of the reaction plane is calculated as the 
weighted sum of the transverse momenta of the measured reaction products with the weights 
chosen to be +(-)1 for products in the forward (backward) c.m. hemisphere~\cite{dani85}.
The obtained corrections are close to one, independent of the specific method,
for the range of higher incident energies ($E > 100A$ MeV) where the directed
flow is large and the reaction plane well defined by the high-multiplicity
distribution of detected particles. At energies below $100$ MeV per nucleon, the
inverse correction factors drop significantly and start to depend on the chosen method. 
The FOPI flow results, as published in Refs.~\cite{andronic01,andronic05} and shown 
in Figs.~\ref{fig:v1corr} and \ref{fig:v2corr}, have been corrected using the
standard method. The midrapidity region of $\pm0.3$ of the scaled c.m. rapidity 
has been excluded to improve the resolution. The corrections used
for the INDRA data have been obtained with a new method adapted to the smaller 
multiplicities and increased emissions of intermediate mass fragments that are
encountered at the lower energies~\cite{andronic06,method_iwm}.

There exists a considerable amount of flow data in the literature with the
potential of being useful for improving our understanding of the development of the
observed collectivity during the high-density phase and its dependence on isospin. 
The FOPI Collaboration has
published a comprehensive report on their measurements of azimuthal asymmetries in 
particle emissions in the regime of 1 GeV per nucleon incident energies~\cite{reisdorf12}.
It includes data for the $^{197}$Au + $^{197}$Au reaction from 90~MeV to 1.5~GeV 
per nucleon but also data for other systems. As an example, the elliptic flow of protons
is shown in Fig.~\ref{fig:fopi_ruzr} for the isotopic pair of mass-symmetric systems 
$^{96}$Zr + $^{96}$Zr and $^{96}$Ru + $^{96}$Ru at the incident energy 1.5~GeV per nucleon.
It is interesting that, even with the precision achieved in this experiment, 
only small differences 
are visible, indicating a minute response to the differences of the corresponding mean fields.
Similar observations have been made by the INDRA/ALADIN collaboration studying several
Xe + Sn reactions at 100 MeV per nucleon with $^{124,129}$Xe projectiles and $^{112,124}$Sn 
targets~\cite{lukasik07}.

\begin{figure}[htb!]
\centering
\includegraphics*[width=1.0\columnwidth]{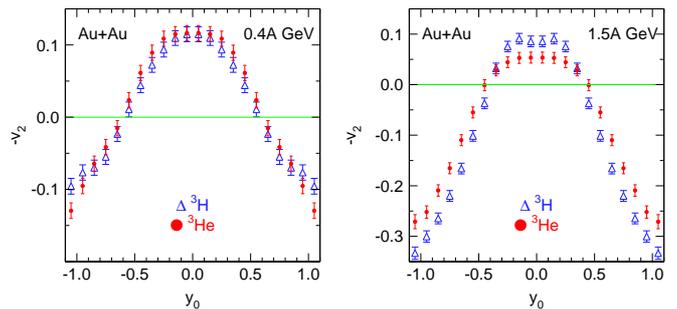}
\vskip -0.1cm
  \caption{Elliptic flow $-v_2 (y_0)$ of $^3$H (open triangles) and $^3$He (filled circles)
in $^{197}$Au + $^{197}$Au collisions for 0.4 (left) and 1.5 GeV per nucleon (right) 
incident energy and centrality $0.25< b_0 < 0.45$. 
Note the inversion of $v_2$ and the definitions of the reduced impact
parameter $b_0 = b/b_{\rm max}$ and of the normalized rapidity $y_0 = y/y_p$ in the c.m. frame
(reprinted from Ref.~\protect\cite{reisdorf12}, Copyright (2012), with permission from 
Elsevier).
}
\label{fig:fopi_3h3he}
\end{figure}    %fig. 9

A complementary type of observable is represented by ratios of mirror nuclei and their 
flow properties. The example of $^3$H and $^3$He elliptic flows in $^{197}$Au + $^{197}$Au 
collisions at 0.4 and 1.5 GeV per nucleon incident energies demonstrates the rich information
provided by exclusive measurements over wide ranges of rapidity and incident 
energies (Fig.~\ref{fig:fopi_3h3he}). 
The values of $v_2$ at midrapidity, their sign changes and their behavior at the spectator 
rapidities are similar at the lower but significantly different at the higher energy. The discussion
of their results by Reisdorf {\em et al.} includes the conjecture that, at the higher energy,
momentum rather than density dependences may be responsible for the increased isotopic 
effect~\cite{reisdorf12}.

\begin{figure}[htb!]
\centering
\includegraphics*[width=1.00\columnwidth]{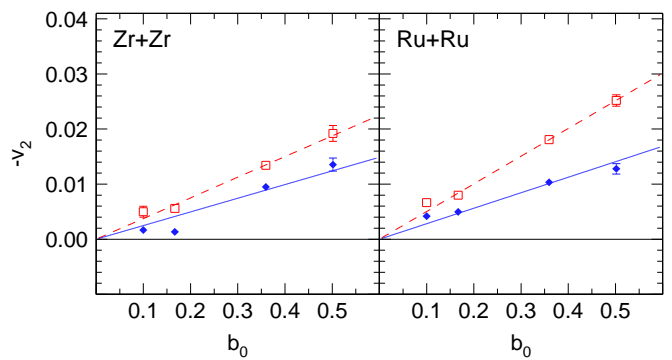}
\vskip -0.1cm
  \caption{Elliptic-flow parameter $-v_2$, integrated over rapidity $-1.8 < y_0 < 0$,
as a function of the reduced impact parameter 
$b_0 = b/b_{\rm max}$ for the systems $^{96}$Zr + $^{96}$Zr (left) 
and $^{96}$Ru + $^{96}$Ru (right) at 1.5 GeV per nucleon.
%($1.8 < y_0 < 0$ and $0.8 < u_{t0} < 4.2$). 
Open squares and dashed lines represent the results for
$\pi^+$ while diamonds and full lines represent $\pi^-$. 
The lines are linear least square fits constrained to $v_2 = 0$ for 
$b_0 = 0$. Note the inversion of $v_2$ and the definition of the reduced 
rapidity $y_0 = y/y_p$ in the c.m. frame
(reprinted from Ref.~\protect\cite{reisdorf07}, Copyright (2007), with permission from 
Elsevier).
}
\label{fig:fopi_pions}
\end{figure}   %fig. 10

The extensive study of pion emission performed by the FOPI Collaboration contains,
besides the pion yield ratios discussed in Section~\ref{sec:probes},  also data 
on pion flows~\cite{reisdorf07}. The example of the impact parameter dependence of the
elliptic flow of pions from $^{96}$Zr + $^{96}$Zr and $^{96}$Ru + $^{96}$Ru collisions
at 1.5 GeV per nucleon is shown in Fig.~\ref{fig:fopi_pions}. Squeeze-out dominates
($v_2 < 0$) and its magnitude is significantly larger for the $\pi^+$ than for the $\pi^-$ case. 
It is also larger for the more proton-rich $^{96}$Ru + $^{96}$Ru system, 
possibly indicating a significant role of Coulomb repulsion.
A detailed discussion of the observed effects is presented, including comparisons with 
the Isospin-QMD (IQMD, Ref.~\cite{hartnack89}) transport model which, however, did not 
fully describe the data. The authors, therefore,
refrained from drawing definite conclusions on the EoS of nuclear matter and suggested
the treatment of the $\Delta$ baryon propagation in the medium as an important topic for further 
studies. It is evident that collective flows may serve as an important complement to yield ratios
in attempts to improve our understanding of the complex pion dynamics.

\section{Transport models and ingredients}
\label{sec:qmd}

To study the sensitivity of the elliptic flow observables to the isospin dependent part
of the equation of state two independent transport models have been employed: 
UrQMD~\cite{qli05,qli06,Li:2006ez} and the T\"{u}bingen version of QMD~\cite{kho92,uma98}. 
They have been upgraded to allow the study of the impact of the symmetry energy on 
observables that can be measured in intermediate-energy collisions of heavy-ions.

\subsection{UrQMD}

This model, originally developed to study particle production at high 
energy~\cite{bass98}, has been adapted to intermediate energy heavy-ion collisions by introducing
a nuclear mean field corresponding to a soft EoS with momentum dependent forces, represented
as two- and three-body Skyrme potentials supplemented by the long range Yukawa and Coulomb 
interactions~\cite{qli09}. A new Pauli-blocking scheme for the suppression of two-body collisions, 
which allows a better description
of experimental observables at lower energies, has also been introduced~\cite{qli11}.
Different options for the dependence on asymmetry were implemented, of which two are used here, 
expressed as a power-law dependence of the potential part of the symmetry energy on the
nuclear density $\rho$ according to
\begin{equation}
E_{\rm sym} = E_{\rm sym}^{\rm pot} + E_{\rm sym}^{\rm kin} 
= 22~{\rm MeV} \cdot (\rho /\rho_0)^{\gamma} + 12~{\rm MeV} \cdot (\rho /\rho_0)^{2/3}
\label{eq:pot_term}
\end{equation}
with $\gamma =0.5$ and $\gamma =1.5$ corresponding to a soft and a stiff density 
dependence (Fig.~\ref{fig:params}).
The kinetic part remains unchanged. 

\begin{figure}[!htb]
 \leavevmode
 \begin{center}
  \includegraphics[width=0.95\columnwidth]{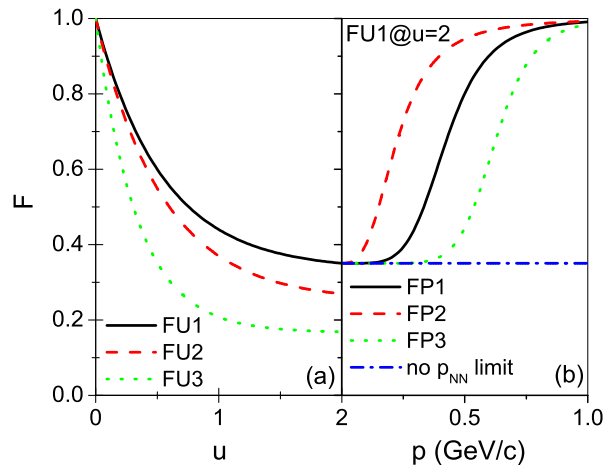}
 \end{center}
\vskip -0.1cm  
  \caption{(a) Correction factor $F$ obtained with the
parameterizations FU1, FU2, and FU3 given in Table~\ref{tabfu} and (b) the
momentum dependence obtained with the four options FP1, FP2,
FP3, and no pNN limit given in Table~\ref{tabfp} for the example of FU1 at
$u = 2$ (reprinted with permission from Ref.~\protect\cite{qli11}; 
Copyright (2011) by the American Physical Society).}

\label{fig:li_nnsections}
\end{figure}    %fig. 11

The issue of the in-medium modification of nucleon-nucleon cross-sections (NNCS) has been
addressed by assuming that the elastic part can be factorized as a product of a medium correction
factor $F(u,\alpha,p)$, with $u = \rho/\rho_0$, and the free elastic NNCS $\sigma_{el}^{free}$: 
\begin{equation}
\sigma_{tot}^*=\sigma_{in}^*+\sigma_{el}^{*}=\sigma_{in}^{free}+F(u,\alpha,p)\,\sigma_{el}^{free}.
\label{nncs}
\end{equation}
%$\sigma_{tot}^*=\sigma_{in}^*+\sigma_{el}^{*}=\sigma_{in}^{free}+F(u,\alpha,p)\,\sigma_{el}^{free}$.
The medium modification factor is proportional to both an isoscalar density effect $F_{u}$ and
an isovector mass-splitting effect $F_\alpha$, which both should be functions
of the relative momentum $p_{NN}$ of the two colliding particles in the NN
center-of-mass system. In Ref.~\cite{Li:2006ez,qli10}, they have been parameterized as
\begin{equation}
F_{\alpha,u}^{\rm p}=\left\{
\begin{array}{l}
f_0 \hspace{3.45cm} p_{NN}>1 {\rm GeV}/c \\
\frac{F_{\alpha,u} -f_0}{1+(p_{NN}/p_0)^\kappa}+f_0 \hspace{1cm}
p_{NN} \leq 1 {\rm GeV}/c.
\end{array}
\right.
\label{fdpup}
\end{equation}
The factor $F_u$ can be expressed as
\begin{equation}
F_u=\lambda+(1-\lambda)\exp[-u/\zeta]. \label{fr}
\end{equation}
Here $\zeta$ and $\lambda$ are parameters which determine the
density dependence of the cross sections while $f_0, p_0$, and $\kappa$ in Eq.~\ref{fdpup}
determine the restoration of the free cross section with increasing relative momentum. 
Several parameter sets, shown in Tables~\ref{tabfu} and \ref{tabfp}, have been selected to
illustrate the resulting modifications of the NNCS (Fig.~\ref{fig:li_nnsections}).

\begin{table}
\begin{center}
\renewcommand{\arraystretch}{1.2}
\begin{tabular}{|l|c|c|}\hline
\bf Set & $\lambda$ & $\zeta$ \\\hline\hline
\tt FU1 \ \ &\  1/3 \  & \ 0.54568  \    \\
\tt FU2 \ \ &\  1/4 \  & \ 0.54568  \      \\
\tt FU3 \ \  &\  1/6 \  & \ 1/3    \    \\ \hline
\end{tabular}
\end{center}
\caption{The three parameter sets FU1, FU2, and FU3 used for the
density-dependent correction factor $F_u$ of elastic NNCS.} \label{tabfu}
\end{table}

\begin{table}
\begin{center}
\renewcommand{\arraystretch}{1.2}
\begin{tabular}{|l|c|c|c|}\hline
\bf Set & $f_0$ & $p_0$ [GeV c$^{-1}$]   & $\kappa$  \\\hline\hline
\tt FP1 & 1   & 0.425 & \ 5  \        \\
\tt FP2 & 1   & 0.225 & 3        \\
\tt FP3 & 1   & 0.625 & 8        \\ \hline
% \tt FP4 & 1.3 1.35? & 0.425 & 4        \\
\tt no $p_{NN}$ limit & $F(u)$ & / &/  \\ \hline
\end{tabular}
\end{center}
\caption{The three parameter sets FP1, FP2, and FP3 used for
describing the momentum dependence of $F_u$. The fourth case,
without a $p_{NN}$ limit, is obtained by setting $f_0$ equal to
$F(u)$ in Eq.\ (\ref{fdpup}).} \label{tabfp}
\end{table}
The reduction of the elastic NNCS as a function of density becomes increasingly more
pronounced as the parameterization is changed from FU1 to FU3. At
the reduced density $u=2$, e.g., the values of FU1, FU2, and FU3 are
0.35, 0.27, and 0.17, respectively. We note that the density
dependence of the FU1 parameterization is in qualitative agreement
with previous work based on the Dirac-Brueckner 
approach~\cite{Li:1993rwa,Li:1993ef,Fuchs:2001fp}. However, in a previous
investigation of the elastic NNCS, based on the effective Lagrangian of
density-dependent relativistic hadron theory in which the $\sigma$,
$\omega$, $\rho$ and $\delta~[a_0(980)]$ mesons are included~\cite{Li:2003vd}, 
it was shown that especially the neutron-proton
cross sections $\sigma_{el,np}^*$ might be largely reduced in the
neutron-rich nuclear medium; the corresponding reduction factor
might be as low as $\sim 0.1$ at $u=2$. Therefore, the other
parameter sets FU2 and FU3 (Table~\ref{tabfu}) are still to be
considered reasonable assumptions.
The importance of the elastic NNCS for describing collective flows in this 
energy range has been confirmed by a study using the recently developed
version of the UrQMD model in which the Skyrme potential energy-density functional 
has been introduced~\cite{wang13}.

\begin{figure}[htb!]
\centering
\includegraphics*[width=80mm]{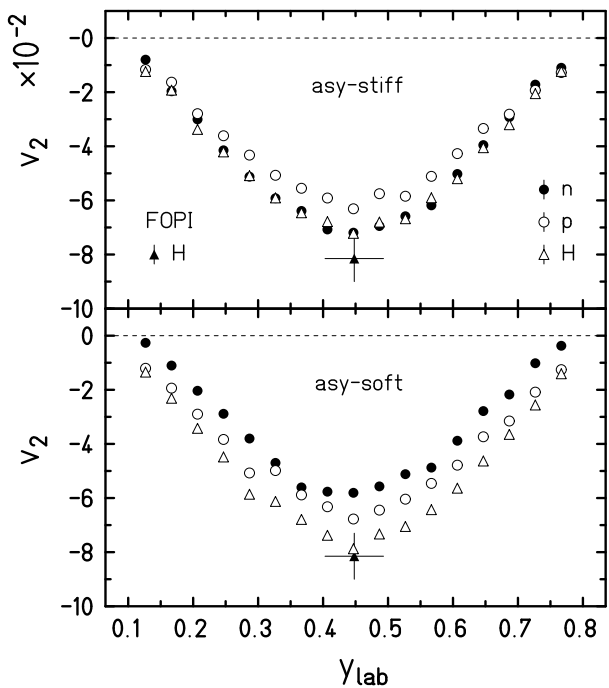}
\vskip -0.1cm
\caption{Elliptic flow parameter $v_2$ for mid-peripheral (5.5 $\le b \le$ 7.5 fm) 
$^{197}$Au + $^{197}$Au 
collisions at 400 MeV per nucleon as calculated with the UrQMD model for neutrons (dots), 
protons (circles), and all hydrogen isotopes ($Z=1$, open triangles), integrated over 
transverse momentum $p_t$, as a function of the laboratory rapidity $y_{\rm lab}$. 
The predictions obtained with a stiff and a soft density dependence of the symmetry term are 
given in the upper and lower panels, respectively. The experimental result from 
Ref.~\protect\cite{andronic05} for $Z=1$ particles at 
mid-rapidity is represented by the filled triangle
% (the horizontal bar represents the experimental rapidity interval)
(reprinted from Ref.~\protect\cite{russotto11}, Copyright (2011), with permission from Elsevier).
}
\label{fig:fig2}
\end{figure}    %fig. 12

The UrQMD transport program is stopped at a collision time
of 150 fm/c at which point a conventional phase-space coalescence
model with two parameters is used to construct clusters. Nucleons
with relative momenta smaller than $P_0$ and relative distances smaller
than $R_0$ are considered as belonging to the same cluster. The values $P_0 = 0.275$~GeV/c
and $R_0 = 3.0$~fm have been adopted as standard parameters.  
With these values the overall dependence of cluster yields on $Z$ is rather well
reproduced but the yields of $Z=2$ particles are under-predicted by a 
factor of 3. The yields of deuterons and tritons in central 
collisions are also underestimated by similar factors~\cite{russotto11}.

According to the UrQMD calculations, one of the most promising probes of the strength 
of the symmetry energy at supra-saturation densities is the difference of the neutron and 
proton (or hydrogen) elliptic flows.
The predictions for the %elliptic flow 
parameter $v_2$ of emitted neutrons, protons, 
and hydrogen isotopes for mid-peripheral $^{197}$Au + $^{197}$Au collisions 
at 400 MeV per nucleon and for the two choices of the density dependence of the symmetry energy, 
labeled asy-stiff ($\gamma = 1.5$) and asy-soft ($\gamma = 0.5$),
are shown in Fig.~\ref{fig:fig2}. They are  displayed as a function of the laboratory 
rapidity $y_{\rm lab}$.
The dominant difference is the significantly larger neutron squeeze-out in the asy-stiff case 
as compared to the asy-soft case. The proton and hydrogen 
flows respond only weakly, and in opposite direction, to the variation of $\gamma$ within the 
chosen stiffness interval.
Relative to each other, the neutron and proton elliptic flows vary on the level of 15\%.
Their absolute magnitude is satisfactorily reproduced with the FP1 parameterization, 
as shown by the comparison with the
FOPI experimental result for $Z=1$ particles at mid-rapidity (Fig.~\ref{fig:fig2}).
Note, however, that this depends sensitively on the chosen parameters as discussed in
Refs.~\cite{russotto11,qli11,wang13}.
%stiff: n .0719 p .0631 soft: n .05807 p .0678

The UrQMD model in the version described here has been employed, with minor changes, to
study the impact of the isovector part of the equation of state of nuclear matter
on a few other observables that can in principle be measured in heavy-ion collisions
or other closely related topics:
the $\Sigma^-/\Sigma^+$ and $\pi^-/\pi^+$ multiplicity ratios~\cite{qli2005}, 
the transverse momentum distribution of the elliptic flow difference~\cite{qli06},
the beam-energy and impact-parameter dependences of the slope parameter of the double 
neutron-to-proton ratio as a function of rapidity~\cite{lili06},
the balance energies of free neutrons~\cite{guo12,wang2012} and the momentum dependence
of the medium modified nucleon-nucleon cross-sections~\cite{qli10,qli11}.

\subsection{T\"{u}bingen QMD}
\label{sec:tueb}
 
To study the impact of the symmetry energy term on elliptic flow observables,
the T\"{u}bingen QMD model~\cite{kho92,uma98} has been expanded by adding an
isospin dependent part to the mean field potential~\cite{cozma11,cozma13}. Two possible
parameterizations have been implemented: a momentum dependent version which has been
developed in Ref.~\cite{das03} starting from the Gogny effective interaction,
\begin{eqnarray}
 U(\rho,\beta,p,\tau,x)&=&A_u(x)\frac{\rho_{\tau'}}{\rho_0}+A_l(x)\frac{\rho_{\tau}}{\rho_0}
+B(\rho/\rho_0)^\sigma(1-x\beta^2) \nonumber\\
&& -8\tau x\frac{B}{\sigma+1}\frac{\rho^{\sigma-1}}{\rho_0^\sigma}\beta\rho_{\tau'} \nonumber\\
&&+\frac{2C_{\tau \tau}}{\rho_0}\,\int d^3 p'\, \frac{f_\tau(\vec{r},\vec{p'})}{1+(\vec{p}-\vec{p}')^2/\Lambda^2}\\
&& +\frac{2C_{\tau \tau'}}{\rho_0}\,\int d^3 p'\, \frac{f_{\tau'}(\vec{r},\vec{p'})}{1+(\vec{p}-\vec{p}')^2/\Lambda^2}\,, \nonumber
\label{eqsympot}
\end{eqnarray}
and a power-law parameterization, momentum independent and similar to that used in 
Ref.~\cite{russotto11}, which has been extended to allow also soft and super-soft 
scenarios~\cite{cozma13}:
\begin{eqnarray}
S(\rho)=\left\{
\begin{array}{l}
 S_0\,(\rho/\rho_0)^\gamma \hspace{0.25cm} \text{-  linear or stiff} \\ 
a+(18.5-a)(\rho/\rho_0)^\gamma  \hspace{0.25cm} \text{-  soft or supersoft}.
\end{array}
\right.
\end{eqnarray}

\noindent
The sets $a = 23.0$~MeV, $\gamma=1.0$ and $a = 31.0$~MeV, $\gamma=2.0$ in the lower 
parameterization reproduce a soft and super-soft density dependence, respectively.
This allowed to test the dependence of the constraints extracted for the symmetry energy on 
its various parameterizations employed in the literature. 

To study the model dependence of elliptic-flow observables, an explicit dependence 
of the microscopic nucleon-nucleon cross-sections on density and isospin 
asymmetry has been introduced. The parameterization of the density dependence of 
elastic proton-proton and neutron-proton cross-sections below the pion production 
threshold obtained by Li and Machleidt~\cite{Li:1993rwa,Li:1993ef} is used. 
The isospin asymmetry dependence is introduced indirectly through a dependence of 
the in-medium nucleon masses on this parameter~\cite{lichen05}. 
This approach is used also above the pion production threshold to simulate the dependence
of nucleon-nucleon cross-sections on both density and isospin asymmetry.

The nucleon optical potential is an important ingredient of transport models which 
still bears some uncertainty on its magnitude.
Its strength can be inferred from different sources: either from first principles~\cite{bal89}
or from the experimental data of proton scattering on Ca and heavier nuclei within 
a relativistic description based on the Dirac-equation
which in turn allowed the extraction of the momentum dependence of the bare 
interaction~\cite{hartnack94}.
The results of the two approaches are somewhat different. The Brueckner-Hartree-Fock approach and its
relativistic counterpart favor a potential that is attractive at all values of the momentum, 
while the relativistic Dirac approach delivers a potential that becomes repulsive above a 
certain momentum threshold, depending on which experimental data sets are considered. 
The Brueckner-Hartree-Fock approach, in addition, predicts an optical potential that 
is almost momentum independent at moderate values of the momentum.

To account for this model dependence, heavy-ion collisions have been simulated by considering 
three different parameterizations of the optical potential. 
The first one stems from the isoscalar part of the Gogny interaction~\cite{das03} while
the last two mimic the parameterizations presented in Ref.~\cite{hartnack94}
\begin{eqnarray}
 V_{opt}^{(MDI)}&=&(C_l+C_u)\,\frac{1}{1+(\vec{p_i}-\vec{p_j})^2/\Lambda^2}\,\frac{\rho_{ij}}{\rho_0} \nonumber
\end{eqnarray}
\vskip -0.5cm
\begin{eqnarray}
V_{opt}^{(HA)}&=&\{V_0+v\,\mathrm{ln}^2[\,a\,(\vec{p_i}-\vec{p_j})^2+1]\}\,\frac{\rho_{ij}}{\rho_0}.
\end{eqnarray}
The values of the parameters can be found in Ref.~\cite{cozma13} for all the cases presented in 
this review.

The QMD program is stopped at $t=60$~fm/c, i.e. also long after the high-density phase of the 
reaction (Fig.~\ref{fig:xu_density}). In order to extract the spectra of free nucleons, 
a density
cutoff at $\rho_0/8$ is applied. The degree of clusterization, as evident from the FOPI 
experimental data, is underestimated by this procedure, however.
A fine tuning of the density cutoff parameter can improve in this respect but the theoretical 
estimates for the elliptic flows of free nucleons depend only mildly on it.

\begin{figure}[!htb]
 \leavevmode
 \begin{center}

 \includegraphics[width=0.9\columnwidth]{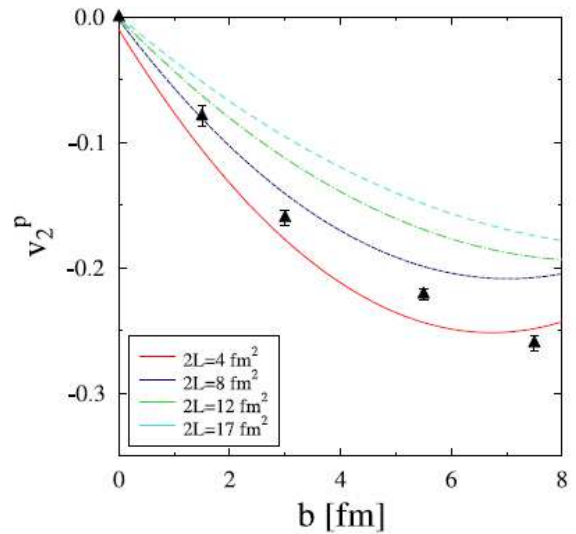}
 \end{center}
\vskip -0.1cm  
  \caption{Elliptic flow of protons as a function of the impact parameter
for various assumptions for the width of the nucleon wave packet.
Note that there are factors of 2 difference between $v_2$ and $L$ as defined in Ref.~\cite{cozma11} 
and here; the ordinate $v_2^p$ of the figure is equal to $2\cdot v_2^p$ as defined 
in Eq.~\ref{eq:defvn} and the values of $2L$ in the legend correspond to the same values of $L$ 
discussed in the text
(reprinted from Ref.~\protect\cite{cozma11}, Copyright (2011), with permission from Elsevier).
}
%obtained with the FOPI acceptance
\label{fig:cozma11}
\end{figure}    %fig. 13

In addition to the already mentioned transport model ingredients, the study performed by 
Cozma~\cite{cozma11} has also addressed the impact of other model parameters on elliptic 
flow observables, in particular those of the nuclear matter compressibility modulus $K$ 
and of the width $L$ of the nucleon wave function. 
The values of these two parameters have been varied around those commonly accepted in 
the literature, $K=210$~MeV and $L=8.66$~fm$^2$~\cite{hartnack98}. It was found that 
the individual elliptic flows of neutrons or protons show a rather sizable sensitivity 
to the parameter 
that is varied. This is presented in Fig.~\ref{fig:cozma11} for the case of $L$ which is varied 
between the extremes $L=4.33$~fm$^2$ and $L=17$~fm$^2$. The former is used when simulating 
collisions of light nuclei while the latter represents a nucleon radius that is 
unrealistically large. The value $L=8.66$~fm$^2$ is the standard choice 
used in simulations of collisions of heavy ions, like $^{197}$Au, as it improves the stability 
of static properties of the respective nuclei.

\begin{figure}[!htb]
 \leavevmode
 \begin{center}
  \includegraphics[width=0.85\columnwidth]{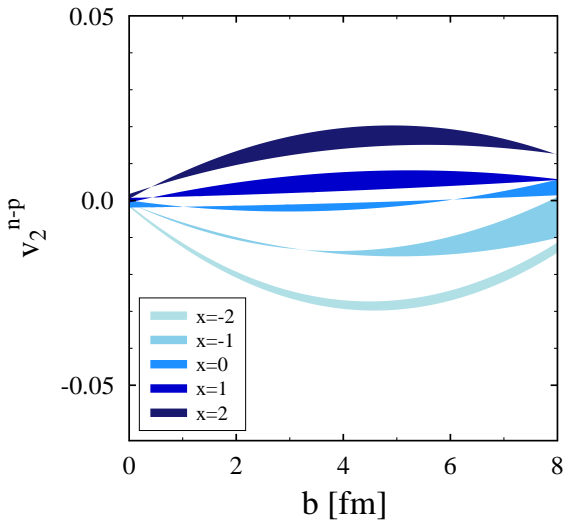}
 \end{center}
\vskip -0.1cm  
  \caption{Sensitivity of the neutron-proton elliptic flow difference 
$v_2^{n-p} = v_2^n - v_2^p$ ($v_2$ as defined in Eq.~\protect\ref{eq:defvn}) 
to the isospin independent EoS and to the choice of 
the symmetry term in the parameterization of Ref.~\protect\cite{lichen05} as
indicated. The widths of the bands represent the uncertainty arising from using either a 
soft ($K = 210$~MeV with momentum dependence) or a hard ($K = 380$~MeV) version for the
isospin-independent part of the EoS. The FOPI acceptance filter has been applied to simulated data
(reprinted from Ref.~\protect\cite{cozma11}, Copyright (2011), with permission from Elsevier).
}
%obtained with the FOPI acceptance
\label{fig:cozma12}
\end{figure}

In contrast, the impact of model parameter changes on the difference of neutron and proton 
elliptic flows was found to be considerably smaller. This is exemplified in Fig.~\ref{fig:cozma12}
by plotting the variation of the neutron-proton elliptic flow difference to changes of the 
compressibility modulus from soft to hard, 
in comparison to variations due to modifications of the stiffness of the symmetry energy from 
super-stiff to super-soft. 
Similar statements hold true also for the case of the width $L$ of the nucleon wave function  
and for various scenarios for the in-medium nucleon-nucleon cross-sections~\cite{cozma11}.

    %fig. 14

The T\"{u}bingen QMD transport model has been successfully applied to the 
description of several heavy-ion related topics in the GeV-per-nucleon regime
of collision energies: dilepton emission~\cite{Shekhter:2003xd,Cozma:2006vp,Santini:2008pk}, 
stiffness of the equation of state of symmetric nuclear matter~\cite{fuchs01} 
and various in-medium effects relevant for the dynamics of heavy-ion 
collisions~\cite{Fuchs:1997we,uma98}.

\section{The FOPI/LAND experiment}

\begin{figure}[htb!]
\centering
\includegraphics*[width=0.85\columnwidth]{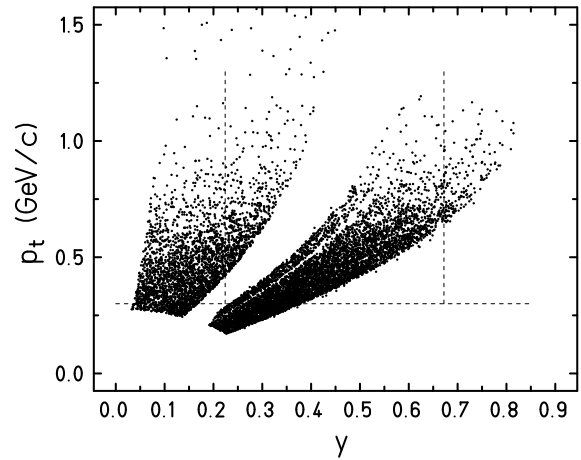}
\vskip -0.1cm
\caption{Scatter plot of 10000 neutron events with energy $E_{\rm lab} \ge 40$~MeV 
in the plane of transverse momentum $p_t$ vs. rapidity $y$ in the laboratory frame 
from a run without shadow bars. 
The quality criteria are the same as used in the flow analysis. Their application produces
the local inefficiency in the forward detector unit caused by individual detector 
modules with reduced performance. 
The dashed lines represent the acceptance cuts $p_t \ge 0.3$~GeV/c and $0.25 \le y/y_p \le 0.75$ 
applied in the analysis
(reprinted from Ref.~\protect\cite{russotto11}, Copyright (2011), with permission from Elsevier).
}
\label{fig:fig0}
\end{figure}

The squeeze-out of neutrons has first been observed by the FOPI/LAND Collaboration who
studied the reaction $^{197}$Au + $^{197}$Au at 400 MeV per nucleon~\cite{leif93}.
The squeeze-out of charged particles reaches its maximum at this energy (Fig.~\ref{fig:v2corr}),
and similarly large anisotropies were observed for neutrons~\cite{lamb94}.
The neutrons have been detected with the Large-Area Neutron Detector, LAND~\cite{LAND}, 
while the FOPI Forward
Wall, covering the forward range of laboratory angles $\theta_{\rm lab} \le 30^{\circ}$ 
with more than 700 plastic scintillator elements, was used to determine the
modulus and azimuthal orientation of the impact parameter. LAND had been divided into two 
separate units which were placed next to each other at sideways positions, providing
the kinematic acceptance shown in Fig.~\ref{fig:fig0}.

  %fig. 15

\begin{figure}[htb!]
\centering
\includegraphics*[width=0.85\columnwidth]{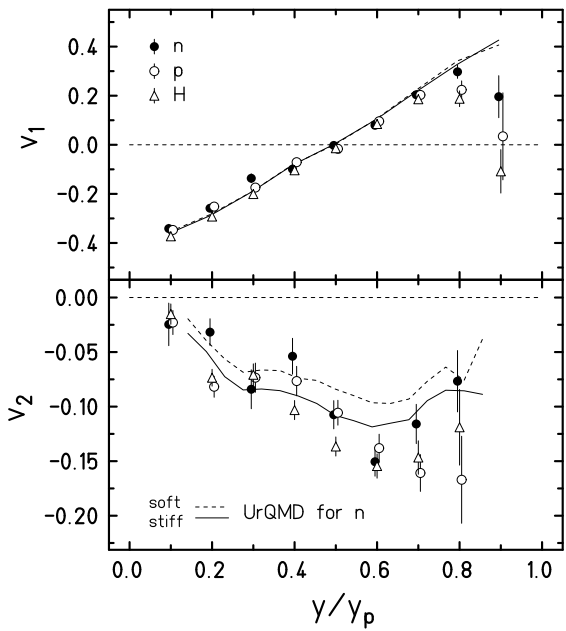}
\vskip -0.1cm
\caption{Measured flow parameters $v_1$ (top) and $v_2$ (bottom)
for mid-peripheral (5.5~$\le b \le$~7.5 fm) $^{197}$Au + $^{197}$Au collisions at 400 MeV per
nucleon for neutrons (dots), protons (circles), and hydrogen isotopes ($Z=1$, open triangles) 
integrated within $0.3 \le p_t/A \le 1.3$~GeV/c per nucleon
as a function of the normalized rapidity $y/y_p$. 
The UrQMD predictions for neutrons are shown for the FP1 parameterization of the in-medium 
cross sections and for a stiff ($\gamma = 1.5$, full lines) and a 
soft ($\gamma = 0.5$, dashed) density dependence of the symmetry term.  
The experimental data have been 
corrected for the dispersion of the reaction plane 
(reprinted from Ref.~\protect\cite{russotto11}, Copyright (2011), with permission from Elsevier).
}
\label{fig:fig3}
\end{figure}

The UrQMD results presented in the previous section provided the motivation for returning 
to the existing data set (Fig.~\ref{fig:fig2}). A reanalysis has been performed which
consisted mainly in choosing 
equal acceptances for neutrons and hydrogen isotopes with regard to 
particle energy, rapidity and transverse momentum (energy and momentum 
per nucleon for deuterons and tritons). The results obtained for a 
mid-peripheral event class are shown in Fig.~\ref{fig:fig3} as a
function of the rapidity $y$, normalized with respect to the projectile rapidity 
$y_p = 0.896$. 
Their asymmetry with respect to mid-rapidity, $y/y_p=0.5$, is caused
by the kinematic acceptance of LAND (Fig.~\ref{fig:fig0}). Its increase in $p_t$ with $y$ and
the decreasing yields at large $p_t$ are responsible for the significant
statistical errors at forward rapidity.
The theoretical predictions have been obtained simulating the LAND acceptance and the 
experimental analysis conditions. The results, shown for neutrons in Fig.~\ref{fig:fig3},
follow qualitatively the experimental data for both, the directed and elliptic flows. However,
in contrast to the elliptic flow, the sensitivity of the directed flow of neutrons to the 
stiffness of the symmetry energy is predicted to be nearly negligible by the UrQMD
(Fig.~\ref{fig:fig3}, top panel).

   %fig. 16

\begin{figure}[htb!]
\centering
   \includegraphics*[width=0.85\columnwidth]{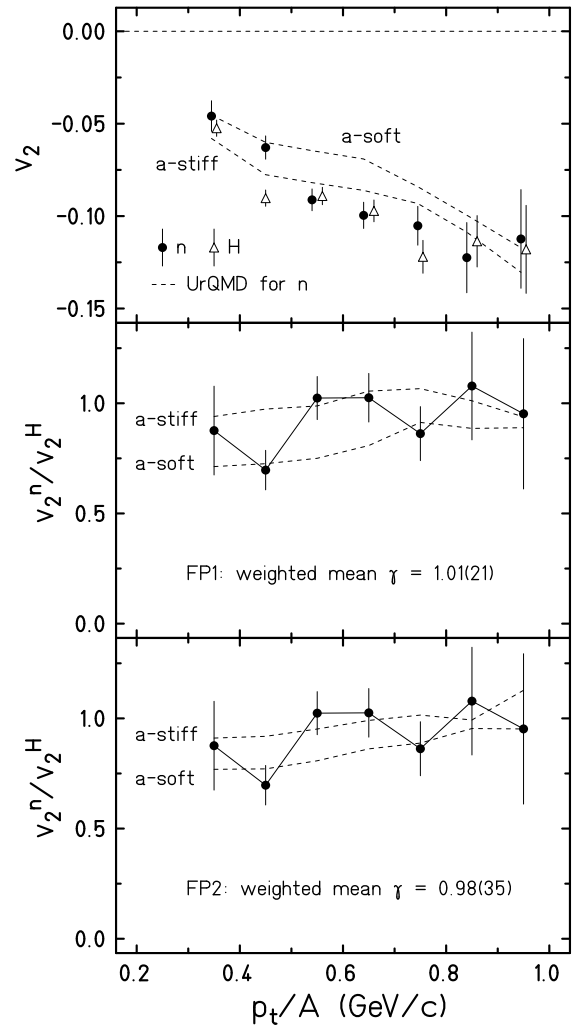}
\vskip -0.1cm
\caption{Elliptic flow parameters $v_2$ for neutrons (dots) and 
hydrogen isotopes (open triangles, top panel) and their ratio (lower panels) for moderately 
central ($b<7.5$ fm) collisions of $^{197}$Au + $^{197}$Au at 400 MeV per nucleon, integrated
within the rapidity interval $0.25 \le y/y_p \le 0.75$, as a function 
of the transverse momentum per nucleon $p_t/A$. The symbols represent the experimental data.
The UrQMD predictions for $\gamma = 1.5$ (a-stiff) and $\gamma = 0.5$ (a-soft)
obtained with the FP1 parameterization for neutrons (top panel) and for 
the ratio (middle panel),
and with the FP2 parameterization for the ratio (bottom panel)
are given by the dashed lines 
(reprinted from Ref.~\protect\cite{russotto11}, Copyright (2011), with permission from Elsevier).
}
\label{fig:fig4}
\end{figure}

The dependence of the elliptic flow parameter $v_2$ on the transverse 
momentum per nucleon, $p_t/A$, is shown in Fig.~\ref{fig:fig4}, upper panel, for the full 
statistics of central and mid-peripheral collisions ($b \le 7.5$~fm) collected in this
experiment. The measured values 
are approximately reproduced by the UrQMD predictions which are significantly 
different for the stiff ($\gamma=1.5$) and soft ($\gamma=0.5$) density dependences. 
For the quantitative comparison, the ratio of the flow parameters of neutrons versus protons 
or versus $Z=1$ particles has been proposed as a useful observable~\cite{russotto11}.  
This choice is expected to minimize systematic effects influencing the collective flows 
of neutrons and charged particles in similar ways. This includes also technical
issues as, e.g., the matching of the impact-parameter intervals used in the 
calculations with the corresponding experimental event groups or the magnitude of the
corrections needed to account for the dispersion of the reaction plane. To test whether this 
kind of insensitivity applies also to model parameters in the isoscalar sector,
the calculations were performed with the two parameterizations FP1 and FP2 of the momentum 
dependence of the elastic nucleon-nucleon cross sections (cf. Fig.~\ref{fig:li_nnsections}).
Their absolute predictions of $v_2$ at mid-rapidity differ by $\approx 40\%$ for this 
reaction~\cite{qli10,qli11}.

   %fig. 17

In Fig.~\ref{fig:fig4} (lower panels), the results for the ratio with respect to the 
total hydrogen yield are shown. 
The calculated ratios exhibit clearly the sensitivity of the elliptic flow to the stiffness 
of the symmetry energy predicted by the UrQMD but depend only weakly on the chosen
parameterization for the in-medium nucleon-nucleon cross section. 
The experimental ratios, even though associated with large errors, 
scatter within the interval given by the two calculations.
Linear interpolations between the predictions, averaged over  
$0.3 < p_t/A \le 1.0$ GeV/c,
yield very similar results $\gamma = 1.01 \pm 0.21$ and $\gamma = 0.98 \pm 0.35$ 
(standard deviations) for the two parameterizations. 
The error is larger for FP2 because the sensitivity is somewhat smaller.

This analysis was repeated in various forms. With the squeeze-out ratios $v_2^n/v_2^p$ of 
neutrons with respect to free protons, similar results were obtained, however with larger errors.
The study of the impact parameter dependence indicated a slightly smaller value 
$\gamma \approx 0.5$ for the mid-peripheral event group, again with larger errors. 
It was also tested to which density region around $\rho_0$ the elliptic-flow
ratios are sensitive, with the result that both,
sub- and supra-saturation densities are probed with this observable~\cite{russotto11}.

In consideration of the apparent systematic and experimental errors, a value
$\gamma = 0.9 \pm 0.4$ has been adopted by the authors as best representing the power-law 
exponent of the potential term resulting from the elliptic-flow analysis. 
It falls slightly below the $\gamma = 1.0$ line shown in Fig.~\ref{fig:params} but, with the
quoted uncertainty, stretches over the interval from $\gamma = 0.5$ halfway up to $\gamma = 1.5$. 
The corresponding slope parameter is $L = 83 \pm 26$~MeV.
According to the UrQMD, the squeeze-out data indicate a moderately soft to linear
behavior of the symmetry energy that is consistent with the density dependence 
deduced from experiments probing nuclear matter near or below saturation.
Comparing with the many-body
theories shown in Fig.~\ref{fig:fuchs06b}, the elliptic-flow result is in good qualitative 
agreement with the range spanned by the DBHF and variational calculations based on realistic
nuclear potentials.

\begin{figure}[!htb]
 \leavevmode
 \begin{center}
  \includegraphics[width=0.9\columnwidth]{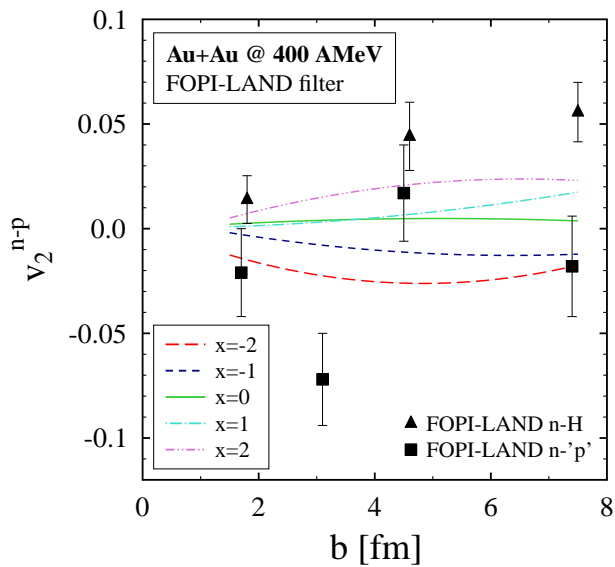}
 \end{center}
\vskip -0.1cm  
  \caption{Comparison between theoretical estimates for the 
neutron-proton elliptic flow difference and the FOPI-LAND experimental data 
for the neutron-proton and neutron-hydrogen elliptic flow differences.
Note that the $v_2$ parameter displayed here is twice larger
than the $v_2$ according to Eq.~\protect\ref{eq:defvn} 
%Note that there is  
because of a factor of 2 difference in the definitions used 
%between $v_2$ as defined 
in Ref.~\cite{cozma11} and here
(reprinted from Ref.~\protect\cite{cozma11}, Copyright (2011), with permission from Elsevier).
}
%obtained with the FOPI acceptance
\label{fig:cozma13}
\end{figure}    %fig. 18

In an independent analysis, Cozma has used data from the same experiment and investigated 
the influence of several parameters on the difference  
between the elliptic flows of protons and neutrons 
using the T\"{u}bingen version of the QMD transport model (Ref.~\cite{cozma11} and 
Section~\ref{sec:tueb}. They included the
parameterization of the isoscalar EoS, the choice of various forms of free or in-medium
nucleon-nucleon cross sections, and model parameters as, e.g., the widths of the wave 
packets representing nucleons. The interaction developed by Das {\it et al.} was used which 
contains an explicit 
momentum dependence of the symmetry energy part~\cite{das03,lichen05}. 
Experimental data for $v_2^p$ presented in Ref.~\cite{lamb94} show a scattered dependence 
on the impact parameter and could not be used to convincingly constrain the stiffness of 
the symmetry energy. 
However, the hydrogen elliptic flow parameter $v_2^H$ published in Ref.~\cite{leif93} 
does not suffer from such a problem and, 
in view of the fact that it represents an upper value (in absolute magnitude) for $v_2^p$, it 
can be used to constrain the high-density dependence of the symmetry energy via the
neutron-hydrogen elliptic flow difference. 
As concluded by Cozma~\cite{cozma11}, an upper limit on the softness of the symmetry energy 
is obtained from
the comparison with the experimental flow data, as can be inferred from Fig.~\ref{fig:cozma13}.

\section{The ASY-EOS experiment}

The ASY-EOS experiment, conducted at the GSI laboratory in 2011, represents an attempt 
to considerably improve the statistical accuracy of the flow parameters for the 
$^{197}$Au + $^{197}$Au system at 400 MeV per nucleon 
but also to complement these measurements with other
observables and data for other systems~\cite{s394,s394_nn2012}. Additional constraints for the
comparison with transport models were considered to be useful for narrowing down the uncertainties
of their predictions as shown in the following section. For this purpose, data were 
collected for two additional systems, the neutron-rich $^{96}$Zr + $^{96}$Zr and 
neutron-poor $^{96}$Ru + $^{96}$Ru pair of mass-symmetric $A=96$ collision systems,
both also at 400 MeV per nucleon incident energy. The FOPI Collaboration has studied
these three reactions in quite some detail, at the present energy 400 MeV per nucleon
but also at other bombarding energies up to 1.5 GeV per nucleon~\cite{reisdorf12}.

The Krak\'{o}w Triple Telescope Array KraTTA~\cite{kratta}
was used to improve the capabilities for measuring charged particle flows under 
the same conditions. With its possibility to identify the masses of light fragments up to
beryllium, the study of isospin effects can be extended to the emission properties of the
isobar pairs $^3$H/$^3$He, $^6$He/$^6$Li, and $^7$Li/$^7$Be. Yield ratios have been suggested 
as useful probes for disentangling ambiguities between the effects of the symmetry energy and 
of the neutron-proton effective-mass splitting in the nuclear 
medium~\cite{ditoro10,feng12,giordano10}.

\begin{figure}[!htb]
 \leavevmode
 \begin{center}
  \includegraphics[width=0.95\columnwidth]{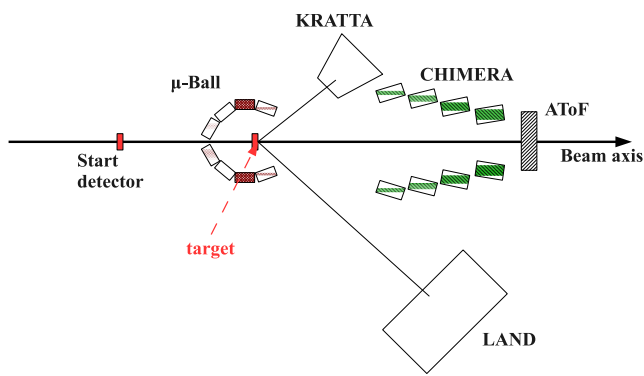}
 \end{center}
\vskip -0.1cm  
  \caption{Schematic view of the experimental setup used in the ASY-EOS
experiment S394 at GSI showing the six main detector systems and their 
positions relative to the beam direction. The dimensions of the symbol 
and the distances are not to scale
(from Ref.~\protect\cite{russotto13}).
%, Copyright (2013), with permission xxx).
}
\label{fig:setup}
\end{figure}   %fig. 19

KraTTA had been specifically designed for the experiment to measure the energy, 
emission angles and isotopic composition of light charged reaction 
products~\cite{kratta}. Its 35 individual triple telescopes were arranged in a 7x5 array
and placed opposite to LAND on the other side of the beam axis at a distance of 
40 cm from the target. Together, they covered 160 msr of solid angle at polar angles 
between $\theta_{\rm lab} = 20^{\circ}$ and 64$^{\circ}$. 

Each KraTTA module consists of two optically 
decoupled CsI(Tl) crystals with thicknesses of 2.5 cm and 12.5 cm and three large area
500 $\mu$m thick PIN photo-diodes. The first photo-diode serves as a $\Delta$E-E 
detector and supplies only an ionization signal. It is followed by the second diode 
mounted at the front entrance of the short CsI(Tl) crystal in a single-chip-telescope 
configuration. It provides both, the intrinsic ionization signal and the light signal
provided by the crystal. The third photo-diode mounted at the back end of the long 
crystal provides the light signal from this third element of the configuration. 
The signals from the photodiodes were processed with custom-made low-noise 
preamplifiers and digitized with wave-form digitizers at a rate of 100 MHz. The separation 
of the ionization and light signals from the second diode and the decomposition of the
fast and slow scintillation components were obtained by individually fitting each of the
recorded signals~\cite{kratta}. 

A schematic view of the experimental set-up is shown in Fig.~\ref{fig:setup}. The beam was
guided in vacuum to about 2~m upstream from the target. A thin plastic foil read by two
photo-multipliers was used to record the projectile arrival times and to serve as a start 
detector for the time-of-flight measurement.
The Large Area Neutron Detector (LAND, Ref.~\cite{LAND}), recently upgraded with new TACQUILA 
GSI-ASIC electronics, was positioned at a laboratory angle near 45$^\circ$ with respect to 
the beam direction, at a distance of about 5~m from the target. Its kinematical acceptance 
was similar to that of the forward LAND subdetector used in the FOPI-LAND experiment 
(cf. Fig.~\ref{fig:fig0}) but slightly larger in rapidity for given transverse momentum 
due to the shorter distance from the target. Also the efficiency was larger because of the 
full 1-m depth that has been used (Fig.~\ref{fig:setup}). The veto-wall of 
plastic scintillators in front of LAND permitted the distinction between neutral and 
charged particles. 

\begin{figure}[!htb]
 \leavevmode
 \begin{center}
  \includegraphics[width=1.00\columnwidth]{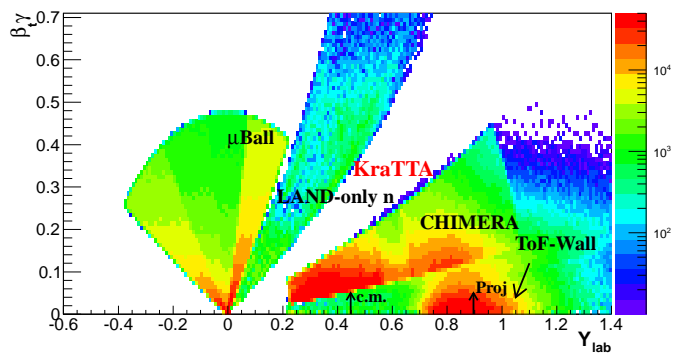}
 \end{center}
\vskip -0.1cm  
  \caption{Invariant hit distribution in the $\gamma \beta_t$ vs. rapidity $y_{\rm lab}$ plane for 
charged particles detected with the three systems $\mu$-Ball, CHIMERA, and ToF-Wall with
full azimuthal coverage and neutrons detected with LAND. The acceptance of KRATTA is 
similar to that of LAND 
(from Ref.~\protect\cite{russotto13}).
%, Copyright (2013), with permission xxx).
}
\label{fig:inv}
\end{figure}    %fig. 20

The determination of the impact parameter and the orientation of the reaction plane 
required several detection devices:\\
i) the ALADIN Time-of-Flight wall~\cite{schuettauf96} was used to detect 
charged particles and fragments in forward direction at polar angles up to
$\theta_{\rm lab} \le 7^{\circ}$. The two walls (front and rear) of about 1-m$^2$
active area consisted of vertical arrays of plastic scintillator slabs, each 2.5-cm
wide, 100-cm long and 1-cm thick, read out by two photo-multipliers at the upper and 
lower ends of the slabs. The front and rear walls were displaced by one half of the slab 
width to gain in horizontal position information. Vertically, a position resolution
of about 10 cm was obtained. Fragment atomic numbers were individually resolved up 
to about $Z=10$. \\ 
ii) 50 thin (between 3.6~mm and 5.6~mm) CsI(Tl) elements of the 
Washington-University $\mu$-ball 
array~\cite{muball}, read out by photo-diodes and arranged in 4 rings to 
cover polar angles between 60$^{\circ}$ and 147$^{\circ}$. The $\mu$-ball elements 
surrounded the target with the aim of measuring the azimuthal distribution of particles
emitted in backward directions in the c.m. system and to discriminate against background 
reactions on non-target material;\\
iii) 352 CsI(Tl) scintillators, 12 cm thick, of the CHIMERA multidetector~\cite{chimera}, 
arranged in 8 rings in 2$\pi$ azimuthal coverage around the beam axis, covered polar 
angles between 7$^{\circ}$ deg and 20$^{\circ}$, measuring the emission of light charged 
particles. In addition, thin (300~$\mu$m) Silicon detectors were placed in front of 32 
(4 by ring) CsI detectors to serve as $\Delta$E detectors. They were used to aid in the 
analysis of the observed pulse shapes of particles stopped in or punching through the 
CsI(Tl) crystals. Particle identification with the CHIMERA elements has been performed 
using pulse-shape analysis based on standard fast-slow techniques. 
Isotopic identification is achieved for p, d, t, and $^{3,4}$He ions stopped in the 
CsI detectors.
\begin{figure}[!htb]
 \leavevmode
 \begin{center}
  \includegraphics[width=0.95\columnwidth]{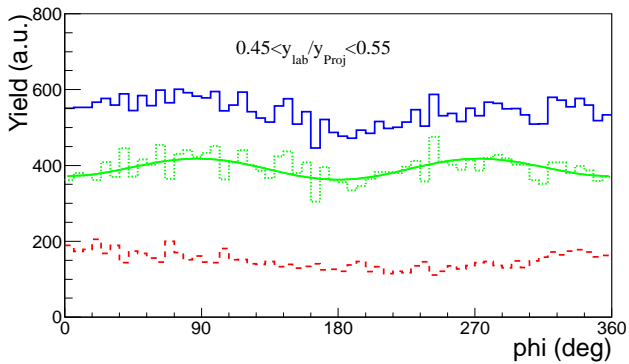}
 \end{center}
\vskip -0.1cm  
  \caption{Azimuthal distributions of neutrons around midrapidity
($0.45 \le y/y_p \le 0.55$),
extracted form the first plane of LAND, with respect
to the reaction plane orientation given by CHIMERA for 
$^{197}$Au + $^{197}$Au collisions at 400 MeV per nucleon. The
lines represent the results of measurements without (top histogram,
blue) and with (bottom histogram, dashed red) the shadow bar inserted.
Their subtraction yields the resulting background-corrected distribution 
given by the dotted green histogram. The full green line represents a fit
according to Eq.~\ref{eq:defvn} 
(from Ref.~\protect\cite{russotto13}).
%, Copyright (2013), with permission xxx).
}
\label{fig:firstv2}
\end{figure}     %fig. 21

With beam intensities of about 10$^5$ particles per second and targets of 1-2\% 
interaction probability, 
about $5\cdot 10^6$ events for each system were collected. Isotopically enriched metallic
$^{96}$Ru and oxide $^{96}$ZrO$_2$ targets with isotopic purities of 96.5\% and 91.4\%,
respecticely, were used. Special runs were performed with 
and without target, in order to measure the background from interaction of projectile 
ions with air, and with iron shadow bars covering the angular acceptance of LAND in
order to measure the background of scattered neutrons. 
The analysis of the collected data is currently 
still in progress and only preliminary results illustrating the overall system performance
are reported here. 

As a global result, the coverage of the various detection systems in the 
transverse-velocity-vs.-rapidity plane is shown in Fig.~\ref{fig:inv}
for the $^{197}$Au + $^{197}$Au system at 400 MeV per nucleon. 
In the case of the $\mu$-ball, without the necessary information on energy and particle 
identity, a uniform kinetic-energy distribution from 0 to 100 MeV has been assumed. 
The populations of two intense regions 
around mid-rapidity and around the projectile rapidity $y_p = 0.896$ are clearly discerned.
Corrections for acceptance or efficiency have not been applied which, e.g., causes the 
discontinuity between the projectile sources as seen with CHIMERA and with the ToF-Wall. 
With common detection thresholds in atomic number $Z$, the discontinuity disappears.

As an example of neutron flow, the azimuthal distributions of neutrons around midrapidity
($0.45 \le y/y_p \le 0.55$), extracted form the first plane of LAND, with respect
to the reaction plane orientation given by CHIMERA alone for $^{197}$Au + $^{197}$Au 
collisions at 400 MeV per nucleon are shown in Fig.~\ref{fig:firstv2}. The
lines represent the results of measurements without and with the shadow bar inserted.
The background intensity is considerable but lacks the squeeze-out pattern seen in the
full and in the background-corrected distributions, the latter being obtained by
subtraction. A fit according to Eq.~\ref{eq:defvn} and 
correction for the attenuation due to the reaction plane dispersion leads 
to a preliminary flow parameter $v_2 = -0.10 \pm 0.01$ for this particular bin in 
rapidity, integrated over impact parameter and over the transverse-momentum range
covered at this rapidity bin (cf. Fig.~\ref{fig:fig0}). It is of the expected magnitude
(Fig.~\ref{fig:fig3}) but interpretations will have to wait for additional
progress of the analysis.

\section{Towards model invariance}

The parameter dependence of the model predictions for the neutron-proton elliptic-flow 
difference (npEFD) and ratio (npEFR) has recently been investigated in 
detail~\cite{cozma13}. The effects of the selected microscopic nucleon-nucleon cross-sections, 
of the compressibility of nuclear matter, of the optical potential, and of the parameterization 
of the symmetry-energy were thoroughly studied. 
The parameterization of the symmetry energy derived from the momentum 
dependent Gogny force 
and a power-law parameterization, as presented in Section~\ref{sec:tueb}, 
were both used in conjunction with the T\"{u}bingen QMD model and the results
were compared with the experimental FOPI/LAND data for $^{197}$Au +  $^{197}$Au 
collisions at 400 MeV per nucleon. The acceptance cuts of the FOPI/LAND experiment were applied. 
It was found that the global variation of 
the general model parameters within commonly accepted limits leads to finite uncertainties 
of typically one unit of the $x$ parameter describing the stiffness of the symmetry energy
in the MDI parameterization~\cite{lipr08,das03,lichen05}.

\begin{figure}[!htb]
 \leavevmode
 \begin{center}
  \includegraphics[width=0.85\columnwidth]{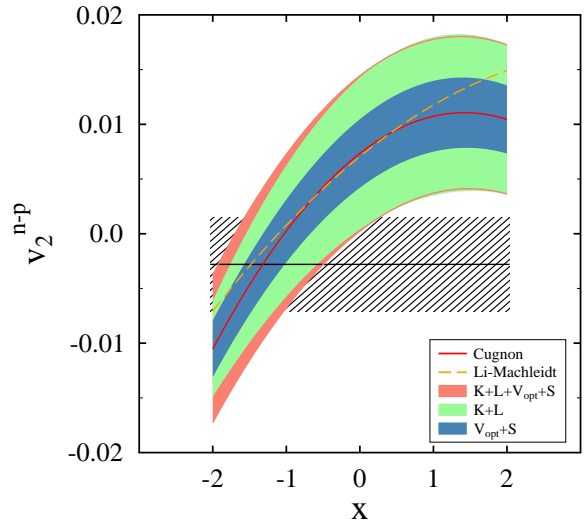}
 \end{center}
\vskip -0.1cm  
  \caption{Model dependence of npEFD in comparison 
with FOPI-LAND experimental data, integrated over impact parameter 
$b < 7.5$~fm, normalized rapidity $0.25 < y/y_p < 0.75$, and transverse momentum
$0.3 < p_t < 1.0$~GeV/c. The sensitivities to the model parameters 
compressibility modulus ($K$), width of the nucleon wave function ($L$), 
optical potential ($V_{opt}$), and parameterization of the symmetry 
energy ($S$) are displayed. The total model dependence is obtained by 
adding in quadrature individual sensitivities.  
%(reprinted from Ref.~\protect\cite{cozma13})
%, Copyright (2013), with permission xxx).
}
\label{fig:cozma22a}
\end{figure}    

\begin{figure}[!htb]
 \leavevmode
 \begin{center}
  \includegraphics[width=0.85\columnwidth]{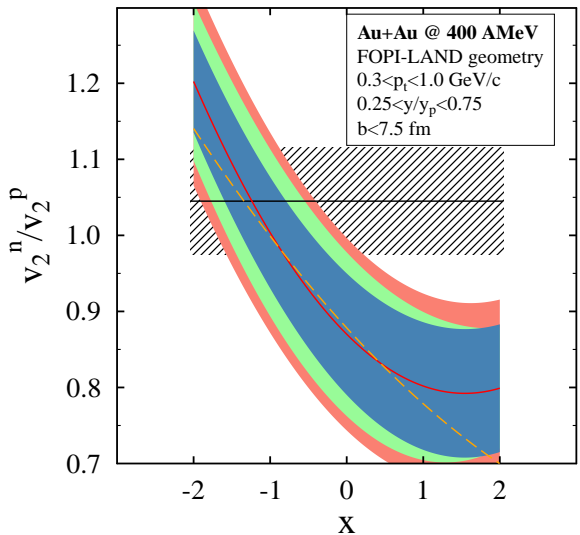}
 \end{center}
\vskip -0.1cm  
  \caption{The same as Fig.~\ref{fig:cozma22a} but for npEFR. 
%(reprinted from Ref.~\protect\cite{cozma13}).
%, Copyright (2013), with permission xxx).
}
\label{fig:cozma22b}
\end{figure}     %fig. 23

The predictions for the npEFD and npEFR are displayed in 
Figs.~\ref{fig:cozma22a},~\ref{fig:cozma22b}, respectively, as a function of the stiffness 
of the asy-EoS. The sensitivity to all model parameters studied in 
Refs.~\cite{cozma11,cozma13} and a 
comparison to the latest impact parameter integrated FOPI/LAND data are included. 
As in Ref.~\cite{cozma13}, the central theoretical results were obtained with a
set of model parameters that best reproduce the experimental values for $v_2^n$ and $v_2^p$
but, in contrast to there, the variations of $K$ and $L$ were not subject to the constraint 
that the elliptic flow values stay within a limit of 25\% with respect to the experimental 
results. Effectively, this has only consequences for the sensitivity to the compressibility 
modulus which has been varied within the range $K=190 \div300$~MeV, giving
rise to departures of $v_2$ from the experimental values by as much as $40-50\%$.

The sensitivities due to the optical potential and symmetry energy parametrizations represent 
averages over the three and respectively two scenarios that were employed here as well as in 
Ref.~\cite{cozma13} for the two model ingredients. The results shown in 
Figs.~\ref{fig:cozma22a},~ \ref{fig:cozma22b} have been obtained by adding in quadrature these 
two uncertainties. Conclusions as, e.g., regarding the sensitivity/insensitivity of the studied 
observables to the momentum dependence of the symmetry energy (as was presented for an asy-soft 
scenario in Ref.~\cite{zhang12}) cannot be directly deduced from these figures. 
Each of the possible combinations of the parameterizations of the optical potential and of the 
symmetry energy usually yields a different outcome in this respect. Nevertheless, it can be 
concluded that
the uncertainties in the optical potential and the momentum dependence/independence of 
the symmetry energy have an important impact on elliptic flow observables like npEFD 
and npEFR. A precise constraining of the high-density symmetry
energy dependence from elliptic flow data will, therefore, require an accurate knowledge 
of the optical potential and the resolution of the problem concerning the momentum
dependence/independence of asy-EoS.

The hatched bands shown in Figs.~\ref{fig:cozma22a},~\ref{fig:cozma22b} represent the
experimental results obtained from the reanalysis of the FOPI-LAND experiment. Their
errors of mainly statistical nature contribute an additional uncertainty, comparable to
the overall parameter dependence. Even with more precise data, a considerable uncertainty 
related to the dependence on model parameters will still remain. It is to be expected, 
however, that the model dependence can be further reduced by explicitly studying trends
of the measured observables as a function of impact parameter, transverse momentum, 
and rapidity. At present, altogether, a moderately stiff, $x=-1.35 \pm 1.25$, symmetry 
energy was extracted from this
study, a result that is slightly stiffer but compatible 
with that of the similar study with the UrQMD transport model 
and a momentum independent power-law parameterization of the symmetry energy.

The uncertainty of $x$ is of statistical nature as far as it originates from the experimental 
errors but mainly systematic regarding the predictions. The widths of the theoretical error bands
are the result of the different scenarios that were considered acceptable. 
Imposing stricter constraints will result in narrower bands. The theoretical errors,
statistical as well as numerical, are about 0.040 for npEFD and 0.035 for npEFR in absolute values,
i.e. about half of the widths of the bands. In the super-soft region, the separation between the 
theoretical and experimental central values is thus of the order of three sigma.

\begin{figure}[!htb]
 \leavevmode
 \begin{center}
  \includegraphics[width=0.85\columnwidth]{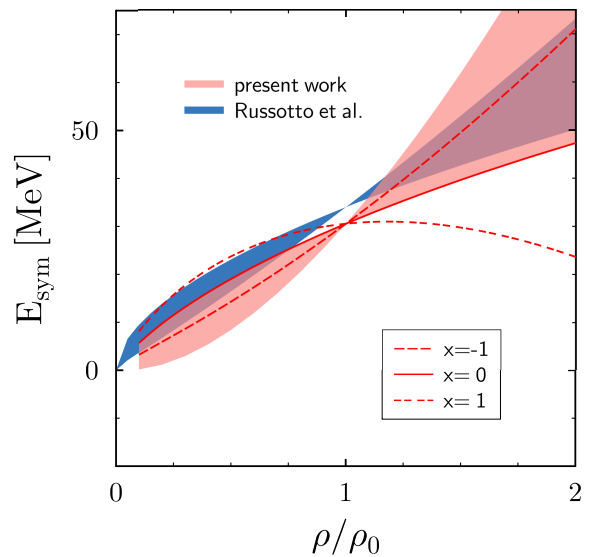}
 \end{center}
\vskip -0.1cm  
  \caption{Constraints on the density dependence of symmetry
energy obtained from comparing theoretical predictions for
npEFD and npEFR to FOPI-LAND experimental data. The
result of Ref.~\cite{russotto11} is also shown together with the Gogny
inspired parameterization of the symmetry energy for three values of the stiffness
parameter: x=-1 (stiff), x=0 and x=1 (soft)
(reprinted with permission from Ref.~\protect\cite{cozma13}; 
Copyright (2013) by the American Physical Society).
}
\label{fig:cozma23}
\end{figure}    %fig. 24

In  Fig.~\ref{fig:cozma23}, the explicit constraints on the density dependence
of the symmetry energy obtained in this study from the comparison of theoretical and 
experimental values 
of npEFD and npEFR are presented. For comparison, the result of Ref.~\cite{russotto11} is added. 
The two studies employ independent flavors of the QMD transport model (T\"{u}bingen
QMD vs. UrQMD) and parameterizations of both, the isoscalar and isovector EoS that differ: 
Gogny inspired vs. Hartnack-Aichelin parameterization~\cite{hartnack94}
of the optical potential (that also differ in their energy dependence) and
Gogny inspired (momentum dependent) vs. power-law parameterization 
(momentum independent) asy-EoS. The constraints on the density dependence of 
the symmetry energy obtained with these different ingredients are in agreement with
each other which contrasts with the current status of the effort to constrain the 
symmetry energy from $\pi^-/\pi^+$ ratios.
By combining the two estimates, a moderately stiff to linear density dependence 
corresponding to a parameterization $x=-1.0 \pm 1.0$ is obtained. It indicates a 
somewhat faster increase of the symmetry energy with density than what is extracted 
from nuclear structure and reactions for sub-saturation densities.

\section{Conclusion and outlook}

According to the predictions of transport models, the relative strengths of
neutron and proton elliptic flows represent an observable sensitive to the symmetry
energy at densities near and above saturation. By forming ratios or differences
of neutron versus proton or neutron versus hydrogen flows, the influence of isoscalar-type
parameters of the model descriptions can be minimized.

The performed comparisons of the existing results of the FOPI/LAND experiment 
for $^{197}$Au + $^{197}$Au reactions at 400 MeV per nucleon with UrQMD transport calculations 
favor a moderately soft to linear symmetry term with a density dependence
of the potential term proportional to $(\rho/\rho_0)^{\gamma}$ with $\gamma = 0.9 \pm 0.4$,
compatible with predictions of ab-initio calculations. A similar result was obtained by studying
the neutron-proton flow differences from the same experiment with the T\"{u}bingen QMD model.
Combining the estimates of both studies leads to the following constraint, obtained from
averaging these results. Expressed in terms of the MDI parameterization, the stiffness of 
the symmetry energy as a function of density is given by $x=-1.0 \pm 1.0$. It corresponds 
to a moderately stiff to linear density dependence and excludes the super-soft and, with a 
lesser degree of confidence, the soft asy-EoS scenarios from the list of possibilities.

Collective flows have proven to be useful probes of nuclear matter properties at high density.
However, the explicit proof that the elliptic-flow ratios are probing the isovector
component of the nuclear mean field at supra-saturation densities is, so far, limited to test 
calculations made with the UrQMD. They indicate that the strength of the symmetry energy 
at densities both, below and above saturation, are essential. It will be useful to study this 
in more detail.

The statistical uncertainty of the existing FOPI/LAND data is larger than the investigated
systematic effects of model parameters and analysis techniques. A significantly improved
result can thus be expected from a new experiment delivering a comprehensive data set
with sufficiently high statistical accuracy. Explicitly testing the predicted dependences on
rapidity and transverse momentum will provide useful constraints for the models. Additional
observables as, e.g., neutron-to-proton or $^3$H-to-$^3$He yield ratios may serve in
resolving ambiguities caused by the effective-mass splitting between neutrons and protons. 
It will be interesting to see first results appearing from the recently completed measurements 
of the ASYEOS Collaboration at the GSI laboratory.

The parameter dependence of the predictions for the differential observables flow ratio and 
flow difference by transport models has been systematically studied with the T\"{u}bingen QMD 
model~\cite{cozma13}. It has been concluded that, while the sensitivity to uncertainties in 
the model parameters is important, the two observables offer the opportunity to extract information
about the symmetry energy above the saturation point. Furthermore, the results of this study
supplemented with those of Russotto et al.~\cite{russotto11} allow one to conclude that constraints 
for the symmetry energy extracted from elliptic flow data are independent of its parameterization, 
suggesting that an almost model independent extraction can be achieved in this case. 
This contrasts with the case of $\pi^-/\pi^+$ ratios where the stiffnesses of asy-EoS extracted 
with different parameterizations or transport models can be extremely different.

New theoretical developments as well as new experimental measurements of meson spectra, 
together with the exploitation of existing data on meson directed and elliptic 
flows~\cite{reisdorf07}, may prove of great importance for approaching a resolution of the 
currently existing pion-ratio problem. It will represent a crucial test of our proper 
understanding of hadronic interactions and their in-medium counterparts in the energy regime
around and below 1 GeV per nucleon. Tighter constraints on the isovector part of the 
equation of state from elliptic flow of nucleons and light clusters appear to be desirable and attainable,
motivating present and future efforts in this direction, both experimentally and theoretically.

Because of the quadratically rising importance of the symmetry energy, the continuation
of this program with systems of larger asymmetry is very promising and
important. The lower luminosities to be expected from the use of secondary beams
and isotopically enriched targets will have to be compensated with efficient 
detector setups. The neutron detector NeuLAND proposed for experiments at FAIR will offer a 
highly improved
detection efficiency for neutrons in the energy range 100 to 400 MeV~\cite{NeuLAND}. 
This will be essential for 
extending the program also to reactions at lower energies for which significant mean-field 
effects are predicted for directed and elliptic flows~\cite{guo12}. 

Stimulating and fruitful discussions with W.~Reisdorf, H.H.~Wolter, and many colleagues within 
the ASY-EOS Collaboration (Ref.~\cite{s394_nn2012}) are gratefully acknowledged.

%\begin{thebibliography}{0}
%\begin{thebibliography}{9}   % Use for  1-9  references


\begin{thebibliography}{99} % Use for 10-99 references
%\itemsep -2pt 

\bibitem{abelev07}   %abelev07,aamodt10,adare12,aad12 
B.I. Abelev {\em  et al.} (STAR Collaboration),    
Phys. Rev. Lett. {\bf 99}, 112301 (2007).

\bibitem{aamodt10} 
K.~Aamodt {\em et al.} (ALICE Collaboration),
Phys. Rev. Lett. {\bf 105}, 252302 (2010). 

\bibitem{adare12}
A.~Adare {\em et al.} (PHENIX Collaboration), 
Phys. Rev. C {\bf 85} 064914 (2012). 

\bibitem{aad12}
G.~Aad {\em et al.} (ATLAS Collaboration),
Phys. Lett. B {\bf 707} 330 (2012).

\bibitem{xu13} 
Jun Xu, Lie-Wen~Chen, Che Ming~Ko, Bao-An~Li, Yu-Gang~Ma,
Phys. Rev. C {\bf 87}, 067601 (2013).

\bibitem{lattprak07}
J.M. Lattimer and M. Prakash, 
Phys. Rep. {\bf 442}, 109 (2007).

\bibitem{lipr08}
%For a recent review, see 
Bao-An Li, Lie-Wen Chen, Che Ming Ko, 
Phys. Rep. {\bf 464}, 113 (2008).

\bibitem{ditoro10}
M. Di Toro, V. Baran, M. Colonna, V. Greco,
J. Phys. G {\bf 37}, 083101 (2010).  

\bibitem{demorest}
P.B.~Demorest, T.~Pennucci, S.M.~Ransom, M.S.E.~Roberts, J.W.T.~Hessels, 
Nature {\bf 467}, 1081 (2010).
%A two-solar-mass neutron star measured using Shapiro delay

\bibitem{fuchs06}
C. Fuchs and H.H. Wolter, 
Eur. Phys. J. A {\bf 30}, 5 (2006).

\bibitem{dani00} 
P.~Danielewicz, 
Nucl. Phys. A {\bf 673}, 375 (2000).
%that the elliptic flow at midrapidity exhibits a particularly strong sensitivity to the meanfield momentum dependence in midperipheral to peripheral collisions. A relatively weak sensitivity was found in these collisions to the incompressibility of nuclear matter.

\bibitem{dani02} 
P.~Danielewicz, R. Lacey, W.G. Lynch, 
Science {\bf 298}, 1592 (2002).

\bibitem{sturm01} 
C. Sturm  {\em et al.}, %[KaoS Coll.],  
Phys. Rev. Lett. {\bf 86}, 39 (2001). 

\bibitem{fuchs01} 
C. Fuchs, A. Faessler, E. Zabrodin, Y.M. Zheng,   
Phys. Rev. Lett. {\bf 86}, 1974 (2001). 
%``Probing the nuclear equation of state by K+ production in heavy ion collisions,''

\bibitem{hartnack06} 
Ch.~Hartnack, H.~Oeschler, J.~Aichelin,
Phys. Rev. Lett. {\bf 96}, 012302 (2006). 

\bibitem{tsang12}
M.B.~Tsang {\em et al.}, 
Phys. Rev. C {\bf 86}, 015803 (2012).

\bibitem{moeller12} 
P.~M\"{o}ller, W.D.~Myers, H.~Sagawa, S.~Yoshida,
Phys. Rev. Lett. {\bf 108}, 052501 (2012).

\bibitem{chen12}
Lie-Wen~Chen, preprint, arXiv:1212.0284[nucl-th] (2012).
 
\bibitem{brown00}
B.A. Brown, 
Phys. Rev. Lett. {\bf 85}, 5296 (2000).

\bibitem{subedi08}
R. Subedi {\em  et al.},
Science {\bf 320}, 1476 (2008). 

\bibitem{xuli10}
Chang Xu and Bao-An Li,
Phys. Rev. C {\bf 81}, 064612 (2010).

\bibitem{steiner12} 
A.W. Steiner and S. Gandolfi, 
Phys. Rev. Lett. {\bf 108}, 081102 (2012). 
%preprint arxiv:1110.4142v2

\bibitem{hebeler10prc}
K.~Hebeler and A.~Schwenk,
Phys. Rev. C {\bf 82}, 014314 (2010).

\bibitem{carb11}
A. Carbone, A. Polls, A. Rios,
Europhys. Lett. {\bf 97}, 22001 (2012).
%preprint arXiv:1111.0797[nucl-th] (2011).

\bibitem{alvioli13}
M.~Alvioli, C.~Ciofi degli Atti, L.P.~Kaptari, C.B.~Mezzetti, H.~Morita, 
Int. J. Mod. Phys. E {\bf 22}, 1330021 (2013).
%preprint arXiv:1306.6235[nucl-th] (2013).

\bibitem{reisdorf97} 
W.~Reisdorf and H.G.~Ritter, 
Annu. Rev. Nucl. Part. Sci. {\bf 47}, 663 (1997).

\bibitem{herrmann99}
N.~Herrmann, J.P. Wessels, T.~Wienold, 
Annu. Rev. Nucl. Part. Sci. {\bf 49}, 581 (1999).

\bibitem{stoecker86} %Theta_flow
H.~St{\"o}cker and W.~Greiner, 
Phys. Rep. {\bf 137}, 277 (1986).

\bibitem{reisdorf12} 
W.~Reisdorf {\em  et al.}, 
Nucl. Phys. A {\bf 876}, 1 (2012).

\bibitem{li02}
Bao-An Li, 
Phys. Rev. Lett. {\bf 88}, 192701 (2002).

\bibitem{lili06}
Q.~Li, Z.~Li, H. St\"{o}cker, 
Phys. Rev. C {\bf 73}, 051601 (2006). 
%nucl-th/0603050.
%double neutron-to-proton ratios

\bibitem{feng12}
Zhao-Qing Feng,
Phys. Lett. B {\bf 707}, 83 (2012). 
%preprint, arXiv:1111.3590[nucl-th] (2011).

\bibitem{famiano06}
M.~Famiano {\em  et al.},  
Phys. Rev. Lett. {\bf 97}, 052701 (2006).

\bibitem{scalone99}
L.~Scalone, M.~Colonna, M.~Di Toro,
Phys. Lett. B {\bf 461}, 9 (1999). 

\bibitem{lisustich01}
Bao-An~Li, A.T.~Sustich, Bin Zhang, 
Phys. Rev. C {\bf 64}, 054604 (2001).

\bibitem{greco03} 
V.~Greco, V.~Baran, M.~Colonna, M.~Di Toro, T.~Gaitanos, H.H.~Wolter, 
Phys. Lett. B {\bf 562}, 215 (2003). 

\bibitem{baran05}
V.~Baran, M. Colonna, V. Greco, M. Di Toro,
Phys. Rep. {\bf 410}, 335 (2005).

\bibitem{gutbrod90} 
H.H.~Gutbrod, K.H.~Kampert, B.~Kolb, A.M.~Poskanzer, H.G.~Ritter, 
R.~Schicker, H.R.~Schmidt,
Phys. Rev. C {\bf 42}, 640 (1990).

\bibitem{russotto11}
P. Russotto {\em  et al.}, 
Phys. Lett. B {\bf 697}, 471 (2011). 

\bibitem{cozma11}
M.D. Cozma, 
Phys. Lett. B {\bf 700}, 139 (2011). 

\bibitem{leif93}
Y. Leifels {\em  et al.}, 
Phys. Rev. Lett. {\bf 71}, 963 (1993).

\bibitem{lamb94}
D. Lambrecht {\em  et al.}, 
Z. Phys. A {\bf 350}, 115 (1994).

\bibitem{LAND}
Th. Blaich {\em  et al.},
Nucl. Instrum. Methods Phys. Res. A {\bf 314}, 136 (1992).

\bibitem{qli05}
%Q. Li {\em  et al.}, 
Q.~Li, Z.~Li, S.~Soff, R.K.~Gupta, M.~Bleicher, H.~St\"{o}cker, 
J. Phys. G {\bf 31}, 1359 (2005).

\bibitem{ferini06}
G.~Ferini, T.~Gaitanos, M.~Colonna, M.~Di~Toro, H.H.~Wolter, 
Phys. Rev. Lett. {\bf 97} 202301 (2006).

\bibitem{xlopez07}
X.~Lopez {\em  et al.}, 
Phys. Rev. C {\bf 75}, 011901(R) (2007).

\bibitem{reisdorf07}
W. Reisdorf {\em  et al.}, 
Nucl. Phys. A {\bf 781}, 459 (2007).

\bibitem{xiao09}
Zhigang~Xiao {\em  et al.}, 
Phys. Rev. Lett. {\bf 102}, 062502 (2009).

\bibitem{feng10}
Zhao-Qing Feng and Gen-Ming Jin, 
Phys. Lett. B {\bf 683}, 140 (2010).

\bibitem{xie13}
W.-J.~Xie, J.~Su, L.~Zhu, F.-S.~Zhang, 
Phys. Lett. B {\bf 718}, 1510 (2013).

\bibitem{wen09}
De-Hua Wen, Bao-An Li, Lie-Wen Chen, 
Phys. Rev. Lett. {\bf 103}, 211102 (2009).

\bibitem{baoan11}
Bao-An Li {\em  et al.},
J. Phys. Conf. Ser. {\bf 312}, 042006 (2011).

\bibitem{das03}
C.B.~Das, S.~Das~Gupta, C.~Gale, Bao-An~Li, 
Phys. Rev. C {\bf 67}, 034611 (2003).

\bibitem{lichen05}
Bao-An Li and Lie-Wen Chen,
Phys. Rev. C {\bf 72}, 064611 (2005).

\bibitem{guo12}
ChenChen~Guo, YongJia~Wang, QingFeng~Li, W.~Trautmann, Ling~Liu, LiJuan~Wu, 
%  Science China Series G} {\bf 55} (2012) 252.
Science China Physics, Mechanics \& Astronomy {\bf 55}, 252 (2012).

\bibitem{ferini05}
G.~Ferini, M.~Colonna, T.~Gaitanos, M.~Di~Toro, 
Nucl. Phys. A {\bf 762}, 147 (2005).

\bibitem{wolter12}
Hermann Wolter,
Proceedings of Science (Bormio2012) 059 (2012).

\bibitem{cozma13}
M.D.~Cozma, Y.~Leifels, W.~Trautmann, Q.~Li, P.~Russotto, 
preprint, arXiv:1305.5417[nucl-th] (2013), accepted for publication in Phys. Rev. C. 

\bibitem{s394}
R.C. Lemmon {\em  et al.}, 
proposal for SIS experiment S394 (2009).

\bibitem{s394_nn2012}
P.~Russotto {\em  et al.}, 
in {\em Proceedings of the 11$^{th}$ International Conference on Nucleus-Nucleus 
Collisions, San Antonio, Texas, USA, 2012}, edited by Bao-An~Li and J.B.~Natowitz,
J. Phys. Conf. Ser. {\bf 420}, 012092 (2013); preprint, arXiv:1209.5961[nucl-ex] (2012).
% Journal of Physics: Conference Series 420 (2013) 012092
% doi:10.1088/1742-6596/420/1/011001

\bibitem{chimera}
A. Pagano {\em  et al.}, 
Nucl. Phys. A {\bf 734}, 504 (2004).

\bibitem{traut12}
W. Trautmann and H.H.~Wolter, 
Int. J. Mod. Phys. E {\bf 21}, 1230003 (2012).

\bibitem{ms} 
W.D.~Myers and W.J.~Swiatecki, 
Nucl. Phys. {\bf 81}, 1 (1966).

\bibitem{baldo04}
M.~Baldo, C.~Maieron, P.~Schuck, X.~Vi\~{n}as,
Nucl. Phys. A {\bf 736}, 241 (2004).

\bibitem{wiringa02}
R.B.~Wiringa and S.C.~Pieper, 
Phys. Rev. Lett. {\bf 89}, 182501 (2002).

\bibitem{zhli06}
Z.H.~Li, U.~Lombardo, H.-J.~Schulze, W.~Zuo, L.W.~Chen, H.R.~Ma, 
Phys. Rev. C {\bf 74}, 047304 (2006).

\bibitem{burgio08}
F.~Burgio,
J. Phys. G {\bf 35}, 014048 (2008).

\bibitem{hebe10}
K.~Hebeler, J.M.~Lattimer, C.J.~Pethick, A.~Schwenk,
Phys. Rev. Lett. {\bf 105}, 161102 (2010).
%EFT up to normal nuclear density; effect of three-body forces

\bibitem{giordano10}
V.~Giordano, M.~Colonna, M.~Di~Toro, V.~Greco, J.~Rizzo,
Phys. Rev. C {\bf 81}, 044611 (2010).

\bibitem{qli06}
Q.~Li, Z.~Li, S.~Soff, M.~Bleicher, H.~St\"{o}cker,
J. Phys. G {\bf 32}, 151 (2006).

\bibitem{Li:2006ez}
Q.~Li, Z.~Li, S.~Soff, M.~Bleicher, H.~St\"{o}cker,
  %``Medium modifications of the nucleon-nucleon elastic cross section in
  %neutron-rich intermediate energy HICs,''
  J.\ Phys.\ G {\bf 32}, 407 (2006).

\bibitem{kho92}
%in-medium effects in the description of heavy-ion collisions with realistic NN interactions
D.T.~Khoa, N.~Ohtsuka, M.A.~Matin, A.~Faessler, S.W.~Huang, E.~Lehmann, R.K.~Puri, 
Nucl. Phys. A {\bf 548}, 102 (1992).

\bibitem{uma98} 
V.S.~Uma~Maheswari, C.~Fuchs, A.~Faessler, L.~Sehn, D.S.~Kosov, Z.~Wang, 
Nucl. Phys. A {\bf 628}, 669 (1998).
%``In-medium dependence and Coulomb effects of the pion production in heavy ion collisions,''

\bibitem{tsang09}
M.B.~Tsang {\em  et al.}, 
Phys. Rev. Lett. {\bf 102}, 122701 (2009).

\bibitem{lattlim12} 
J.M. Lattimer and Y. Lim,
ApJ. {\bf 771}, 51 (2013).
%preprint, arXiv:1203.4286[nucl-th] (2012).

\bibitem{steiner10}   %steiner10,steiner13
A.W.~Steiner, J.M.~Lattimer, E.F.~Brown,
ApJ. {\bf 722}, 33 (2010).

\bibitem{steiner13} 
A.W.~Steiner, J.M.~Lattimer, E.F.~Brown,
ApJ. {\bf 765}, L5 (2013).
%review in arXiv:1005.0811v2
%ApJ Lett. 765, L5, 2013
%THE NEUTRON STAR MASS RADIUS RELATION AND THE EQUATION OF STATE OF DENSE MATTER
 
\bibitem{baoan12} 
Bao-An~Li, Lie-Wen~Chen, F.J.~Fattoyev, W.G.~Newton, Chang~Xu,
preprint, arXiv:1212.1178[nucl-th] (2012).

\bibitem{vidana09} 
I.~Vida\~{n}a, C.~Provid\^{e}ncia, A.~Polls, A.~Rios,
Phys. Rev. C {\bf 80}, 045806 (2009). 

\bibitem{abrah12}
S.~Abrahamyan {\em et al.}, 
Phys. Rev. Lett. {\bf 108}, 112502 (2012).

\bibitem{horo12}
C.J. Horowitz {\em  et al.},
Phys. Rev. C {\bf 85}, 032501(R) (2012). 
%preprint, arXiv:1202.1468[nucl-ex] (2012).
%PREX Phys. Rev. Lett. 108, 112502 (2012) 

\bibitem{bertsch87}
G.F.~Bertsch, W.G.~Lynch, M.B.~Tsang, 
Phys. Lett. B {\bf 189}, 384 (1987).

\bibitem{lemmon99} 
R.C.~Lemmon {\em et al.}, 
Phys. Lett. B {\bf 446}, 197 (1999).
%Direct observation of the inversion of flow 
%doi:10.1016/S0370-2693(98)01545-7       

\bibitem{westfall98} 
G.D.~Westfall, 
Nucl. Phys. A {\bf 630}, 27c (1998).

\bibitem{westfall93} 
G.D.~Westfall {\em et al.}, 
Phys. Rev. Lett. {\bf 71}, 1986 (1993).

\bibitem{pak97}      %pak97,pak97a
R.~Pak {\em et al.}, 
Phys. Rev. Lett. {\bf 78}, 1022 (1997). 

\bibitem{pak97a} 
R.~Pak {\em et al.}, 
Phys. Rev. Lett. {\bf 78}, 1026 (1997).

\bibitem{zhang06}
Yingxun Zhang and Zhuxia Li, 
Phys. Rev. C {\bf 74}, 014602 (2006).

\bibitem{kohley10}
Z.~Kohley {\em et al.}, 
Phys. Rev. C {\bf 82}, 064601 (2010).

\bibitem{kohley12}
Z.~Kohley {\em et al.}, 
Phys. Rev. C {\bf 85}, 064605 (2012).

\bibitem{qli11}
Qingfeng Li, C.~Shen, C.~Guo, Y.~Wang, Z.~Li, J.~Lukasik, W.~Trautmann, 
Phys. Rev. C {\bf 83}, 044617 (2011).

\bibitem{li_npa02}
Bao-An Li, 
Nucl. Phys. A {\bf 708}, 365 (2002).

\bibitem{yong06}
Gao-Chan Yong, Bao-An Li, Lie-Wen Chen,
Phys. Rev. C {\bf 74}, 064617 (2006).

\bibitem{voloshin96}
S. Voloshin and Y. Zhang, 
Z. Phys. C {\bf 70}, 665 (1996).
%different definition with x and y coeff.

\bibitem{ollitrault97} 
J.-Y. Ollitrault, preprint, arXiv:nucl-ex/9711003 (1997).
%correct definition with Sum2vn...

\bibitem{poskanzer98} 
A.M. Poskanzer and S.A. Voloshin, 
Phys. Rev. C {\bf 58}, 1671 (1998).
%correct definition with Sum2vn...

\bibitem{andronic01} 
A.~Andronic {\em  et al.}, 
Phys. Rev. C {\bf 64}, 041604 (2001).

\bibitem{andronic06}
A. Andronic, J.~{\L}ukasik, W.~Reisdorf, W.~Trautmann,
Eur. Phys. J. A {\bf 30}, 31 (2006).

\bibitem{lukasik05} 
J. {\L}ukasik {\em  et al.}, 
Phys. Lett. B {\bf 608}, 223 (2005).

\bibitem{andronic05} 
A. Andronic {\em et al.}, 
Phys. Lett. B {\bf 612}, 173 (2005).

\bibitem{pinkenburg99}
C. Pinkenburg {\em  et al.}, 
Phys. Rev. Lett. {\bf 83}, 1295 (1999). 

\bibitem{bmunzinger98} 
P.~Braun-Munzinger and J.~Stachel, 
Nucl. Phys. A {\bf 638}, 3c (1998).

\bibitem{tsang93}
M.B.~Tsang {\em  et al.}, 
Phys. Rev. C {\bf 47}, 2717 (1993). 

\bibitem{fries08}
R. Fries, V. Greco, P. Sorensen, 
Annu. Rev. Nucl. Part. Sci. {\bf 58}, 177 (2008). 

\bibitem{snell11}
R. Snellings, 
{\it New J. Phys.} {\bf 13}, 055008 (2011).

\bibitem{dani85} 
P.~Danielewicz and G.~Odyniec, 
Phys. Lett. B {\bf 157}, 146 (1985).

\bibitem{method_iwm} 
J.~{\L}ukasik and W.~Trautmann, 
in {\em Proceedings of the IWM2005 International Workshop on Multifragmentation and Related Topics, 
Catania, Italy, 2005}, edited by R. Bougault {\em et al.}, Conf. Proc. Vol. {\bf 91} 
(Italian Physical Society, Bologna 2006) p. 387; preprint, arXiv:nucl-ex/0603028 (2006). 

\bibitem{lukasik07} 
J.~{\L}ukasik and W.~Trautmann, 
preprint, arXiv:0708.2821[nucl-ex] (2007).

\bibitem{hartnack89}
C.~Hartnack, Li.~Zhuxia, L.~Neise, G.~Peilert, A.~Rosenhauer, H.~Sorge, J.~Aichelin,
H.~St\"{o}cker, W.~Greiner,
Nucl. Phys. A {\bf 495}, 303c (1989).

\bibitem{bass98}
S.A. Bass {\em  et al.},
Progr. Part. Nucl. Phys. {\bf 41}, 225 (1998).

\bibitem{qli09}
Q.~Li and M.~Bleicher, 
J. Phys. G {\bf 36}, 015111 (2009).

\bibitem{qli10}
Qingfeng Li, Caiwan~Shen, M.~Di~Toro,
Mod. Phys. Lett. A {\bf 25}, 669 (2010).
%``Probing the momentum dependence of medium modifications of the nucleon-nucleon 
%elastic cross sections,''

\bibitem{Li:1993rwa}
  G.Q.~Li and R.~Machleidt,
  %``Microscopic calculation of in-medium nucleon-nucleon cross-sections,''
  Phys.\ Rev.\  C {\bf 48}, 1702 (1993).

\bibitem{Li:1993ef}
  G.Q.~Li and R.~Machleidt,
  %``Microscopic calculation of in-medium proton proton cross-sections,''
  Phys.\ Rev.\  C {\bf 49}, 566 (1994).

\bibitem{Fuchs:2001fp}
  C.~Fuchs, A.~Faessler, M.~El-Shabshiry,
  %``Off shell behaviour of the in medium nucleon-nucleon cross section,''
  Phys.\ Rev.\  C {\bf 64}, 024003 (2001).

\bibitem{Li:2003vd}
  Q.~Li, Z.~Li, E.~Zhao,
  %``Density and temperature dependence of nucleon-nucleon elastic cross
  %section,''
  Phys.\ Rev.\  C {\bf 69}, 017601 (2004).

\bibitem{wang13}
Y.~Wang, C.~Guo, Q.~Li, H.~Zhang, Z.~Li, W.~Trautmann,
preprint, arXiv:1305.4730[nucl-th] (2013).

\bibitem{qli2005}
  Q.~Li, Z.~Li, E.~Zhao, R.K.~Gupta,
  %``Sigma-/Sigma+ ratio as a candidate for probing the density dependence of the symmetry potential at high nuclear densities,''
  Phys.\ Rev.\ C {\bf 71}, 054907 (2005).

\bibitem{wang2012}
 Y.~Wang, C.~Guo, Q.~Li, H.~Zhang,
  %``The effect of symmetry potential on the balance energy of light particles emitted from mass symmetric heavy-ion collisions with isotopes, isobars and isotones,''
  Sci.\ China Phys.\ Mech.\ Astron.\  {\bf 55}, 2407 (2012).

\bibitem{bal89}
M.~Baldo, I.~Bombaci, G.~Giansiracusa, U.~Lombardo, 
Phys. Rev. C {\bf 40}, R491 (1989).

\bibitem{hartnack94}
C.~Hartnack and J.~Aichelin, 
Phys. Rev. C {\bf 49}, 2801 (1994).

\bibitem{hartnack98}
the notation for $L$ follows the convention used in
Ch.~Hartnack, R.K.~Puri, J.~Aichelin, J.~Konopka, S.A.~Bass, H.~St\"{o}cker, W.~Greiner,
Eur. Phys. J. A {\bf 1}, 151 (1998).

\bibitem{Shekhter:2003xd} 
  K.~Shekhter, C.~Fuchs, A.~Faessler, M.~Krivoruchenko, B.~Martemyanov,
  %``Dilepton production in heavy ion collisions at intermediate energies,''
  Phys.\ Rev.\ C {\bf 68}, 014904 (2003).

\bibitem{Santini:2008pk} 
  E.~Santini, M.D.~Cozma, A.~Faessler, C.~Fuchs, M.I.~Krivoruchenko, B.~Martemyanov,
  %``Dilepton production in heavy-ion collisions with in-medium spectral functions of vector mesons,''
  Phys.\ Rev.\ C {\bf 78}, 034910 (2008).

\bibitem{Cozma:2006vp} 
  M.D.~Cozma, C.~Fuchs, E.~Santini, A.~Fassler,
  %``Dilepton production at HADES: Theoretical predictions,''
  Phys.\ Lett.\ B {\bf 640}, 170 (2006).

\bibitem{Fuchs:1997we} 
  C.~Fuchs, P.~Essler, T.~Gaitanos, H.H.~Wolter,
  %``Temperature and thermodynamic instabilities in heavy ion collisions,''
  Nucl.\ Phys.\ A {\bf 626}, 987 (1997).

\bibitem{kratta}
J. {\L}ukasik {\em  et al.}, 
Nucl. Instrum. Methods Phys. Res. A {\bf 709}, 120 (2013).

\bibitem{russotto13}
P.~Russotto {\em  et al.}, 
in {\em Proceedings of the International Nuclear Physics Conference INPC2013, Firenze, Italy, 2013}, 
to appear in EPJ Web of Conferences (2013).

\bibitem{schuettauf96} % Tof-Wall
A. Sch\"{u}ttauf {\em et al.}, 
Nucl. Phys. A {\bf 607}, 457 (1996).

\bibitem{muball}
D.G.~Sarantites, P.-F.~Hua, M.~Devlin, L.G.~Sobotka, J.~Elson, J.T.~Hood,
D.R.~LaFosse, J.E.~Sarantites, M.R.~Maier, 
Nucl. Instr. and Meth. A {\bf 381}, 418 (1996).

\bibitem{zhang12}
L.~Zhang, Y.~Gao, Y.~Du, G.H.~Zuo, G.C.~Yong, 
Eur. Phys. J. A {\bf 48}, 30 (2012).
%arXiv:1203.1724[nucl-th].

\bibitem{NeuLAND}
NeuLAND Technical Design Report, submitted to FAIR (2011).




\end{thebibliography}
\end{document}